\documentclass{aastex62}
\usepackage[utf8]{inputenc}

\usepackage{subfiles}
\usepackage{color}
\usepackage{longtable}
\usepackage{multirow}
\graphicspath{{./}{figures/}}

\accepted{30 May 2019}
\submitjournal{ApJ}

\shorttitle{Potential Targets for Ariel}
\shortauthors{Edwards et al.}

\begin{document}
\title{An Updated Study of Potential Targets for Ariel}
\correspondingauthor{Billy Edwards}
\email{billy.edwards.16@ucl.ac.uk}

\author[0000-0002-5494-3237]{Billy Edwards}
\affil{Department of Physics and Astronomy, University College London, Gower Street, London, WC1E 6BT, UK}

\author[0000-0002-9007-9802]{Lorenzo Mugnai}
\affil{Dipartimento di Fisica, La Sapienza Universita di Roma, Piazzale Aldo Moro 2, 00185 Roma, Italy}

\author[0000-0001-6058-6654]{Giovanna Tinetti}
\affil{Department of Physics and Astronomy, University College London, Gower Street, London, WC1E 6BT, UK}

\author[0000-0002-3242-8154]{Enzo Pascale}
\affil{Dipartimento di Fisica, La Sapienza Universita di Roma, Piazzale Aldo Moro 2, 00185 Roma, Italy}
\affil{School of Physics and Astronomy, Cardiff University, Queens Buildings, The Parade, Cardiff, CF24 3AA, UK}

\author{Subhajit Sarkar}
\affil{School of Physics and Astronomy, Cardiff University, Queens Buildings, The Parade, Cardiff, CF24 3AA, UK}

\begin{abstract}

Ariel has been selected as ESA's M4 mission for launch in 2028 and is designed for the characterisation of a large and diverse population of exoplanetary atmospheres to provide insights into planetary formation and evolution within our Galaxy. Here we present a study of Ariel's capability to observe currently-known exoplanets and predicted TESS discoveries.

We use the Ariel Radiometric model (ArielRad) to simulate the instrument performance and find that $\sim$2000 of these planets have atmospheric signals which could be characterised by Ariel. This list of potential planets contains a diverse range of planetary and stellar parameters. From these we select an example Mission Reference Sample (MRS), comprised of 1000 diverse planets to be completed within the primary mission life, which is consistent with previous studies.

We also explore the mission capability to perform an in-depth survey into the atmospheres of smaller planets, which may be enriched or secondary. Earth-sized planets and Super-Earths with atmospheres heavier than H/He will be more challenging to observe spectroscopically. However, by studying the time required to observe $\sim$110 Earth-sized/Super-Earths, we find that Ariel could have substantial capability for providing in-depth observations of smaller planets.

Trade-offs between the number and type of planets observed will form a key part of the selection process and this list of planets will continually evolve with new exoplanet discoveries replacing predicted detections. The Ariel target list will be constantly updated and the MRS re-selected to ensure maximum diversity in the population of planets studied during the primary mission life.

\end{abstract}

\section{Introduction}
\label{sec:intro}

As of March 2019, nearly 4000 exoplanets have been discovered (around 3000 of which transit their stars) as well as 2900 \textit{Kepler} and \textit{K2} candidates yet to be confirmed as planets. On top of this, in the next few years Gaia is anticipated to discover up to ten thousand Jupiter-sized planets \citep{sozzetti,perryman} while the Transiting Exoplanet Survey (TESS, \cite{ricker}) is expected to detect thousands of transiting planets of Earth size or larger \citep{sullivan,barclay,huang}. Additionally, space surveys CHEOPS (CHaracterising ExOPlanet Satellite, \cite{broeg}) and PLATO (PLAnetary Transits and Oscillations of stars, \cite{rauer}), along with ground-based surveys like NGTS (Next-Generation Transit Survey, \cite{wheatley}), WASP (Wide Angle Search for Planets, \cite{pollacco}), HATNet (Hungarian-made Automated Telescope Network, \cite{bakos}), HATSouth (Hungarian-made Automated Telescope Network-South, \cite{bakos_south}), MEarth \citep{nutzman}, TRAPPIST (Transiting Planets and Planetesimals Small Telescope, \cite{jehin}) and KPS (Kourovka Planet Search, \cite{burdanov}), will lead to many more transiting exoplanet detections as well as further characterisation of planetary parameters.

Although many planets have been detected and it is thought that planets are common in our Galaxy (e.g. \cite{howard,batalha,cassan,dressing,wright_jup}), our current knowledge of their atmospheric, thermal and compositional characteristics is still very limited. Space telescopes such as \textit{Hubble} and \textit{Spitzer}, as well as some ground-based observatories, have provided constraints on these properties for a limited number of targets and, in some cases, have identified the key molecules present in their atmospheres whilst also detecting the presence of clouds and probing the thermal structure (e.g. \cite{brogi,majeau,stevenson,sing,fu,tsiaras_30planets,zhang,pinhas}). However, currently available space-based datasets have been achieved with instruments that are not specifically designed for exoplanet science. Therefore, the data obtained is inhibited due to a narrow wavelength coverage and, where observations are taken over a wider spectral range, these observations are usually not simultaneous, potentially injecting an extra source of systematic noise. Additionally, being general observatories, the time allocated to exoplanet science does not fully meet the need of the community. Additionally, the current observatories are not dedicated missions and thus the time allocated to exoplanet science is limited. Hence, the breadth and quality of currently available data is limited by the absence of a dedicated space-based exoplanet spectroscopy mission and thus progress in this area has been slower than desired. A dedicated mission would also provide a heterogeneous dataset, with a consistent pipeline and an well-defined target selection strategy, maximising the scientific yield.

Ariel has been selected as the next ESA medium-class science mission and is due for launch in 2028. During its 4-year mission, Ariel aims to observe $\sim$1000 exoplanets ranging from Jupiters and Neptunes down to Super-Earth size in the visible and the infrared with its meter-class telescope. The analysis of Ariel spectra and photometric data will deliver a homogeneous catalogue of planetary spectra which will allow for the extraction of the chemical fingerprints of gases and condensates in the planets’ atmospheres, including the elemental composition for the most favourable targets. It will also enable the study of thermal and scattering properties of the atmosphere as the planet orbits around the star. A basic summary of the Ariel's instrumentation is given in Table \ref{Instrument}. For more detail on the Ariel design see \citep{tinetti_ariel}.

\begin{table}[h!]
\centering
\begin{tabular}{lll}
\hline
Instrument Name & Wavelength Range [$\mu$m] & Resolution \\ \hline
VISPhot & 0.5 - 0.6 & \\
FGS 1 & 0.6 - 0.81 & Photometric Bands\\
FGS 2 & 0.81 - 1.1 &  \\ \hline
NIRSpec & 1.1 - 1.95 & 20 \\ \hline
AIRS Ch0 & 1.95 - 3.9 & 100 \\ \hline
AIRS Ch1 & 3.9 - 7.8 & 30 \\ \hline
\end{tabular}
\label{Instrument}
\caption{Wavelength ranges and spectral resolutions of Ariel's instrumentation}
\end{table}

Ariel will enable the simultaneous study of exoplanets at multiple wavelengths through transit, eclipse and phase-curve observations (see e.g. \cite{tinetti_spec} for an overview of the information content of these techniques). During transit, stellar light can be observed passing through the terminator region of the planet (transmission spectroscopy). Similarly, when the star eclipses the planet (i.e. the planet passes behind its host star in our line of sight) the flux difference resulting from the planet's day-side emission or reflection (emission or reflection spectroscopy) can be measured. Phase-curves are observed by monitoring the star-planet system over a large portion of the planet’s orbit. Here, we focus on transit and eclipse observations as these will be the main science observations. Additional science time to be dedicated to phase-curves is currently under study.

During Phase A, a study of Ariel's capabilities to observe known and predicted planets was conducted and a Mission Reference Sample (i.e a list of exoplanets to be observed during the primary mission life) of $\sim$1000 potential targets was created \citep{zingales}. Here an updated review of the performance of Ariel's instrumentation to observe currently-known planets and potential future detections by TESS is undertaken. According to a recent study by \cite{barclay}, TESS is anticipated to detect over 4500 planets around bright stars and nearly 10,000 giant planets around fainter stars. The predicted TESS discoveries are incorporated into our analysis to test Ariel's capabilities. The list of known and predicted exoplanets are analysed using ArielRad \citep{mugnai}, a new Ariel simulator which is more suitable to capture the details and updates of the Ariel's design as considered in Phase B (see Section \ref{ArielRad}). ArielRad includes greater margins on the instrument noise and an additional noise floor of 20ppm than the previously used ESA radiometric model \citep{puig}. This exercise will be regularly repeated to incorporate new discoveries and validate that the mission's science goals can be achieved as the instrumentation evolves in Phase B.

Finally we focus part of our simulations and discussion on smaller planets, to refine some of the science objectives considered in Phase A for the mission and address new science questions emerging from the recent discoveries, e.g.  the ``Fulton gap'' \citep{fulton}. 

\section{Creation of a Catalogue of Exoplanets}
\label{catalogue}
\subsection{Known Exoplanets}

Exoplanetary data was downloaded from NASA's Exoplanet Archive in order to account for all confirmed planets before being filtered such that only transiting planets were considered. The database was last accessed on 26$^{th}$ February 2019. However, the major exoplanet catalogues are sometimes incomplete and thus an effort has been made here to combine them (for a review of the current state of exoplanet catalogues see \cite{christiansen}). 

Hence, the data was verified, and in some cases gaps filled, utilising the Open Exoplanet Catalogue \citep{rein}, exoplanet.eu \citep{schneider} and TEPCat \citep{southworth}. Planets not included in the NASA Exoplanet Archive were not added to the analysis to ensure that only confirmed planets were utilised. As of March 2019, 3022 planets within the NASA Exoplanet Archive were sufficiently characterised for inclusion in this analysis.

Unknown parameters were inferred based on the following assumptions:

\begin{itemize}
   \item If the inclination is known, the impact parameter is calculated from:
    \begin{equation}
   b = \frac{ acos(i) }{R_*}
   \end{equation}
   \item Else, it was assumed that \textit{b} = 0.5 (i.e. the midpoint of the equator and limb of the star)
   \item Planetary effective temperature (\textit{T$_p$}) is estimated from: 
   \begin{equation}
   T_p =  T_* \left (\frac{ \sqrt[]{1- A} R_* }{2a\epsilon} \right )^{1/2}
   \end{equation}
   
   where a greenhouse effect of $\epsilon = 0.8 $ and a planetary albedo of \textit{A} = 0.3 (\textit{T$_P$} $<$  700K) or \textit{A} = 0.1 (\textit{T$_p$} $>$ 700K) are assumed \citep{tessenyi,seager}
   
   \item Planetary mass (\textit{M$_p$}) was estimated utilising Forecaster \citep{chen}. 
   \item Atmospheric molecular mass was assumed to be 2.3
   
\end{itemize}

\subsection{Future Planet Discoveries}

TESS and other surveys are predicted to discover thousands of planets around bright stars. In the first two years of operation, TESS is anticipated to detect over 4500 planets around bright stars and more than 10,000 giant planets around fainter stars \citep{barclay}. Here, these predicted TESS discoveries around brighter stars are incorporated into the analysis to highlight Ariel's capabilities to study anticipated future discoveries. The MAST archive\footnote{https://archive.stsci.edu} has been utilised to obtain stellar parameters for these planets by cross-referencing the Gaia catalogue. The first planets from TESS have begun to be discovered (e.g. \cite{huang}) but these have not been included in this work to avoid overlap with the predicted yield. The known and predicted exoplanets were compiled into a single dataset ($\sim$7000 planets) which has been used to analyse Ariel's capabilities and provide an indicative look at the number and type of planets Ariel could observe.

Potential discoveries by other surveys (PLATO, SPECULOOS etc.) have not been included in this analysis as thus far predictions for these surveys just resulted in an estimate of the number of expected detections, but no specific target coordinates and characteristics have been released. When such information becomes available, predicted/real detections from these surveys will be incorporated into this analysis. In any case, these surveys are expected to find thousands of planets which could be suitable for study with Ariel, enhancing the population of planets from which the final target list (Mission Reference Sample) is selected. Hence, although these predicted yields have not been included, planets found by these surveys will be added to the sample as they are detected.

\section{Creating a List of Potential Targets}
\label{MRS}

\subsection{ESA Radiometric Model}

During Phase A, the ESA radiometric model \citep{puig} was utilised to assess the  duration and type of observations needed to meet the mission requirements. Although the NIRSpec instrument will also be used for spectroscopy, the mission requirements are baselined on the AIRS channels, as these bands are typically the most demanding. The ESA Radiometric Model calculates the signal and noise contributions for exoplanet spectroscopic observations \citep{puig,sarkar}. This model simulates observational and instrumentation effects, utilising target characteristics to assess whether emission or transmission spectroscopy is preferable and to estimate the required number of observations to achieve a desired resolving power and signal to noise ratio (SNR). The ESA radiometric model requires the host star temperature to be in the range 3070-7200K. The Mission Reference Sample during Phase A was obtained using this model \citep{zingales}.

\subsection{Ariel Radiometric Model}
\label{ArielRad}

The ESA radiometric model assumes the systematic noise does not vary from target to target. The Ariel Radiometric model (ArielRad, \cite{mugnai}) has been developed to provide a comprehensive model of the instrument performance. While the ESA radiometric model assumes a constant instrument noise, ArielRad provides systematic noise on a case by case basis. The Ariel team has validated ArielRad against the ESA radiometric model and ExoSim \citep{sarkar} by running the simulators with the same instrument noise characteristics. ArielRad includes greater margins on the instrument noise and a noise floor of 20ppm.

We use the ArielRad simulator to provide realistic noise models for all planets within the catalogue described in Section \ref{catalogue}. These noise models are used to create a new list of potential targets, based on the expected performance from ArielRad. The FGS signal requirements for accurate pointing are now accounted for as these are not included in the ESA radiometric model but are key for target selection. In the ESA radiometric model, simulations were restricted to planets orbiting stars with temperatures in the range 3070-7200K due to the stellar spectral energy distributions (SEDs) used. For ArielRad, this range is expanded to include early type stars and M-dwarfs such as Trappist-1 by using a broader range of SEDs from the Phoenix atmospheric models increasing the diversity of input catalogue.

\subsection{The 3 Tier Approach}

Planning of observations with Ariel is based around a tiered approach and Table \ref{tiers} describes the requirements on each tier. As envisaged in Phase A, a survey tier aims to observe 1000 planets with low resolution spectroscopy to produce a statistically viable dataset of a diverse range of exoplanetary atmospheres. Tier 1 observations will help refine orbital and planetary parameters and constrain (or remove) degeneracies in the interpretation of mass-radius diagrams. Additionally, it will offer the opportunity to generate colour/colour and colour/magnitude diagrams and investigate what fraction of planets have a transparent atmosphere, are partially clouded or are completely overcast.

\begin{table}[h]
\centering
\begin{tabular}{llll}
\hline
Instrument Name & Tier 1 & Tier 2 & Tier 3 \\ \hline
NIRSpec & R$\sim$1 & R$\sim$10 & R$\sim$20 \\
AIRS Ch0 & R$\sim$3 & R$\sim$50 & R$\sim$100 \\
AIRS Ch1 & R$\sim$1 & R$\sim$10 & R$\sim$30 \\ \hline
\end{tabular}
\caption{Resolution of final dataset across each instrument in each tier}
\label{tiers}
\end{table}

\begin{figure}
    \centering
    \includegraphics[width = 1\textwidth]{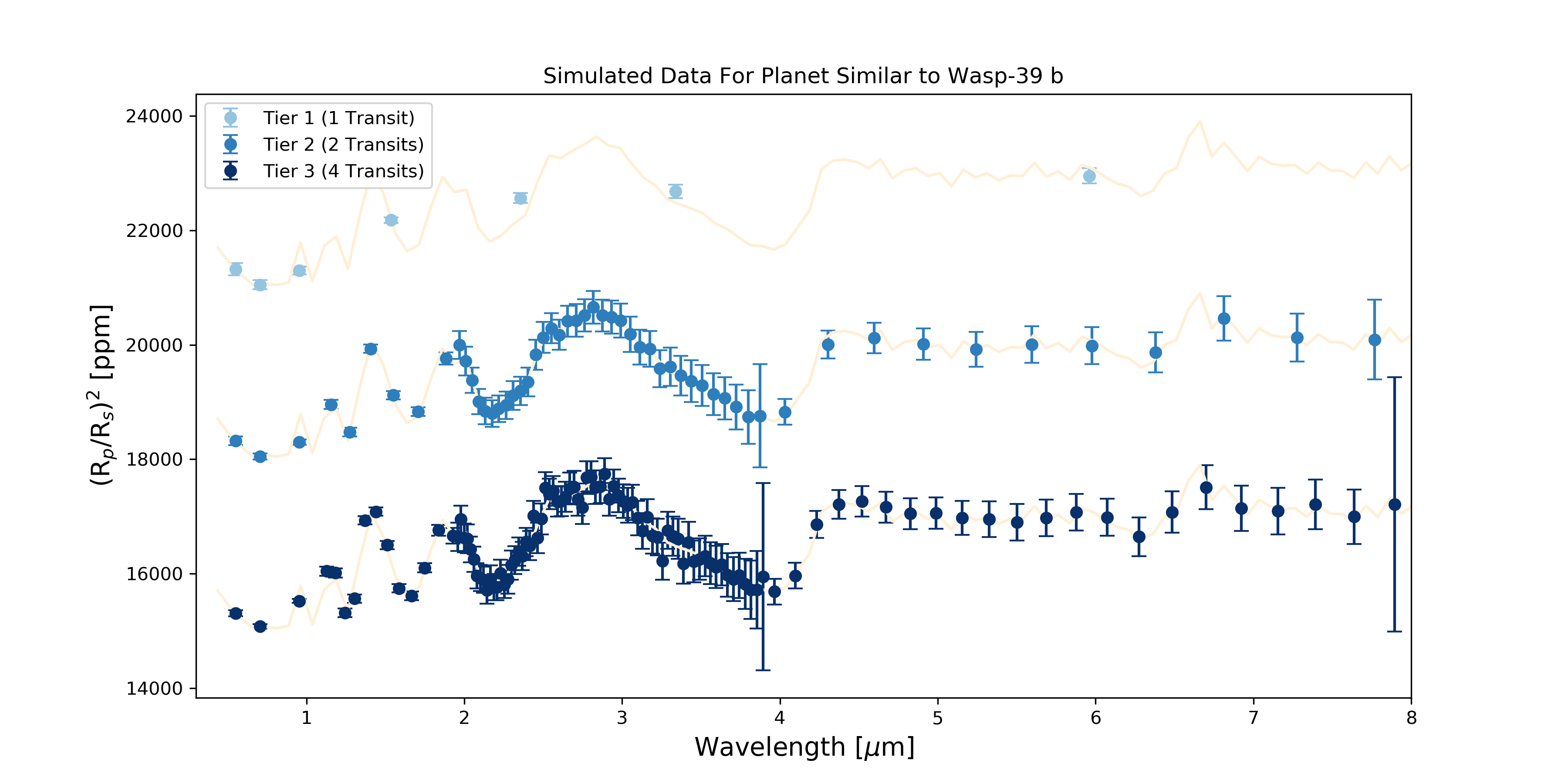}
    \caption{Simulated data for a planet similar to Wasp-39 b in each Tier. The atmosphere has been modelled in chemical equilibrium with solar metallicity and C/O = 0.5. The error bars are calculated using ArielRad and the spectra are offset for clarity. The larger errors at the red end of AIRS Channels 0 and 1 are due to a reduced sensitivity caused by optical filter cut-off, and detector sensitivity, respectively. This will however be mitigated by the cross-channel spectral overlap of the baseline design which is expected to reduce the error bars at the transition between channels 0 and 1.}
    \label{example tiers}
\end{figure}

From this initial survey of planets, around half will be selected for spectroscopic follow-up: Tier 2 spectroscopic measurements are crucial for uncovering atmospheric structure and composition. Additionally, Tier 2 observations are critical to search for potential correlations between atmospheric chemistry and basic parameters such as planetary size, density, temperature, stellar type and metallicity. Tier 3 will consist of repeated observations of select group of benchmark planets ($\sim$50-100) around bright stars which can be observed at high resolution within a small number of transits or eclipses to provide a very detailed knowledge of the planetary chemistry and dynamics (see \cite{tinetti_ariel} for an in-depth description of the tiering system and the mission science questions and requirements). Figure \ref{example tiers} shows simulated observations in each tier for a planet with parameters similar to Wasp-39 b. The addition of a Tier 4 -- including phase-curves and an ad-hoc observational strategy for targets of interest which do not fit into the tier system -- has been recently discussed in the Ariel team.

\subsection{A List of Potential Targets for Ariel}
\label{list targets}
\begin{figure}[h!]
    \centering
    \includegraphics[width = 0.37\textwidth]{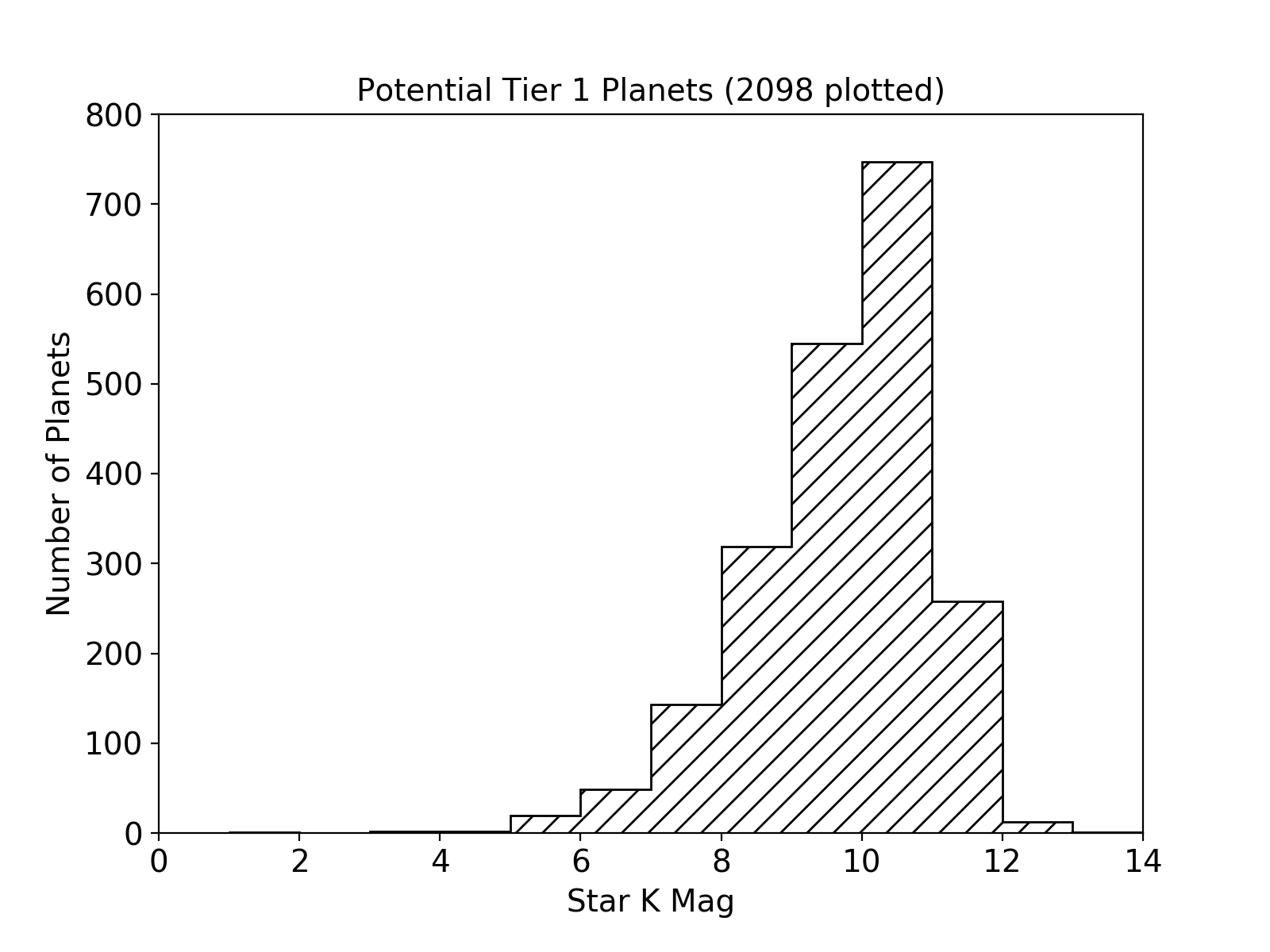}
    \includegraphics[width = 0.37\textwidth]{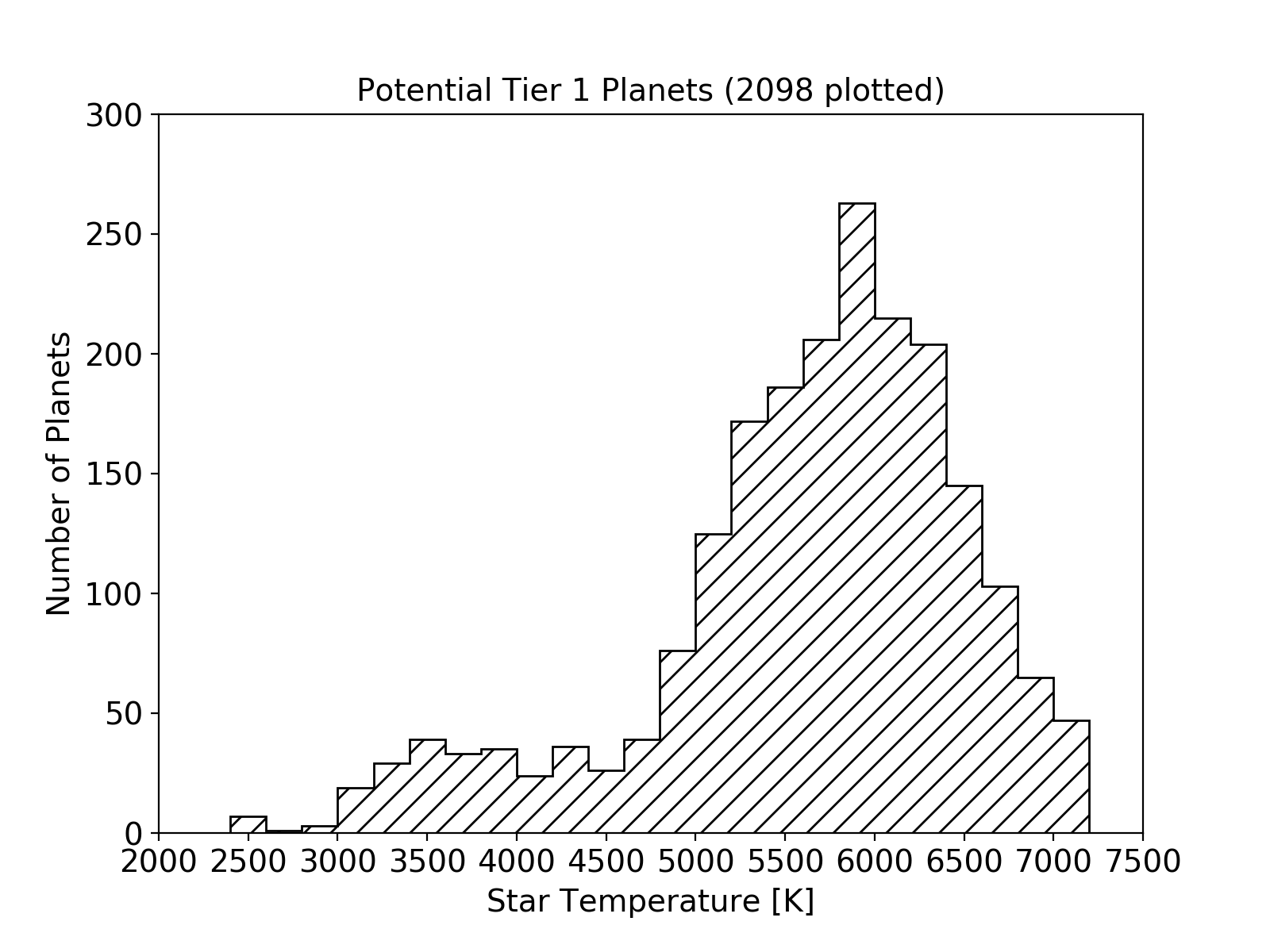}
    \includegraphics[width = 0.37\textwidth]{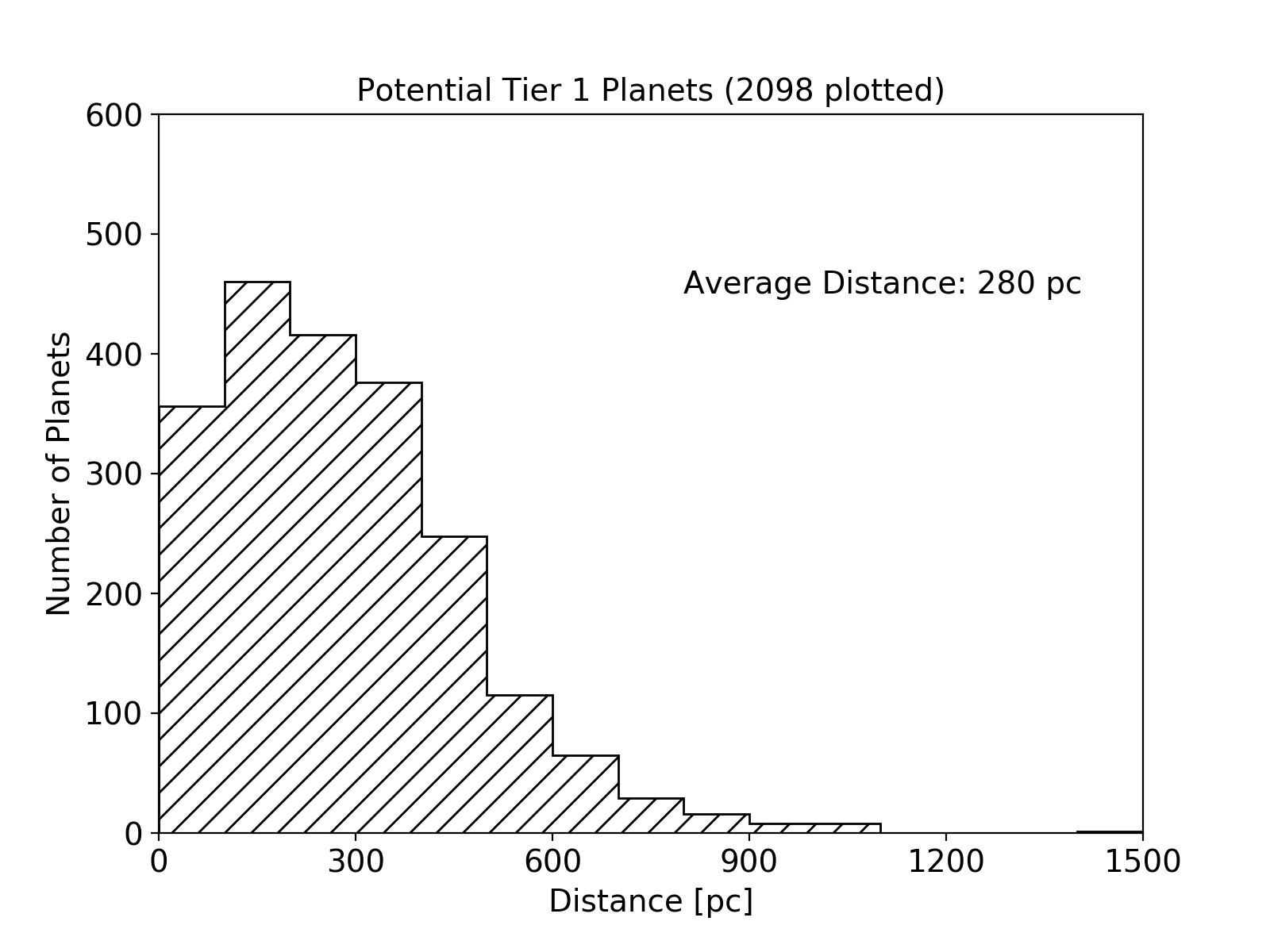}
    \includegraphics[width = 0.37\textwidth]{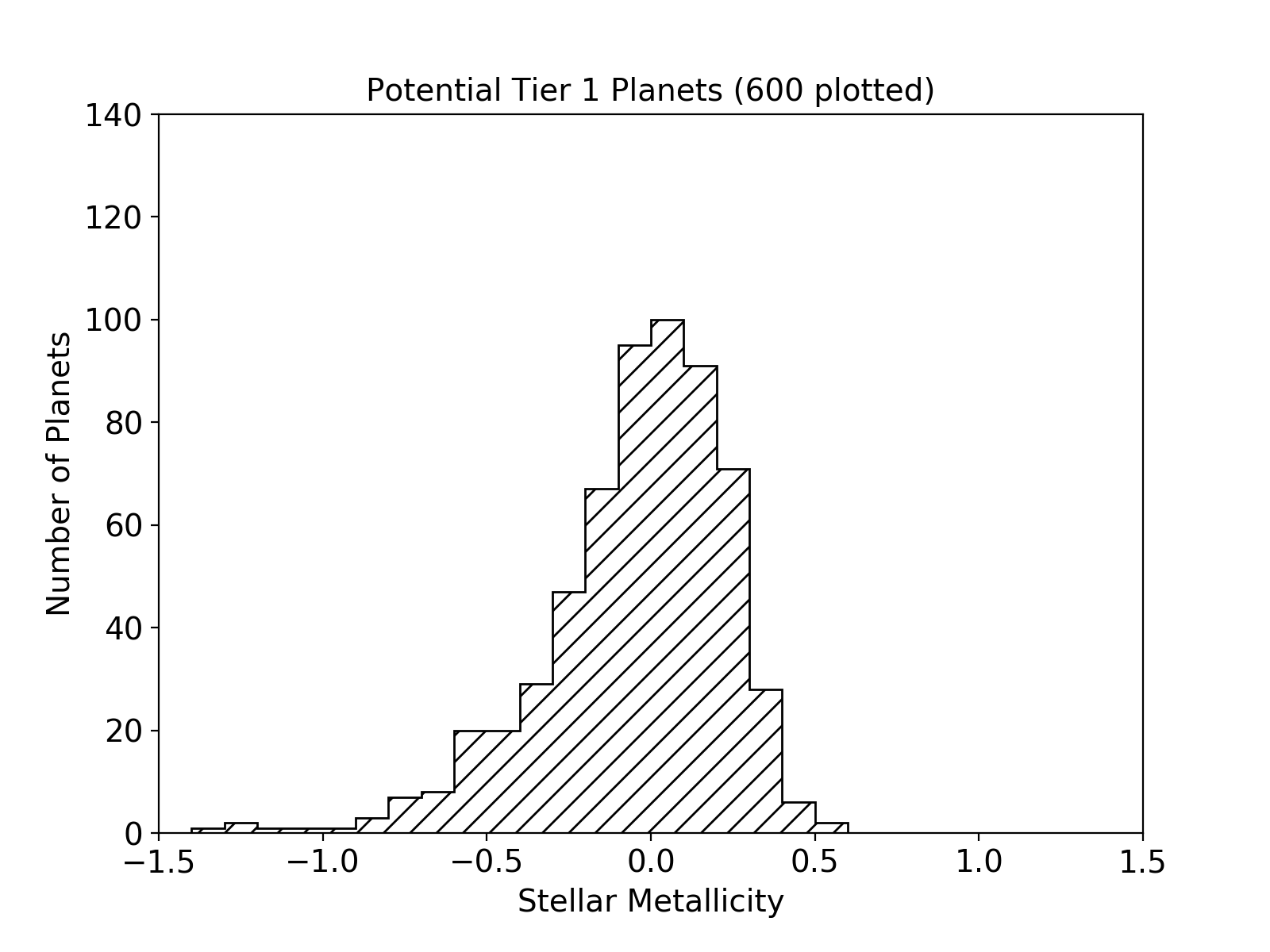}
    \caption{Histograms of the properties of the stellar hosts within the potential Ariel Tier 1 Catalogue. Metallicities were not available for all host stars.}
    \label{stellar para}
\end{figure}

\begin{figure}[h!]
    \centering
    \includegraphics[width = 0.37\textwidth]{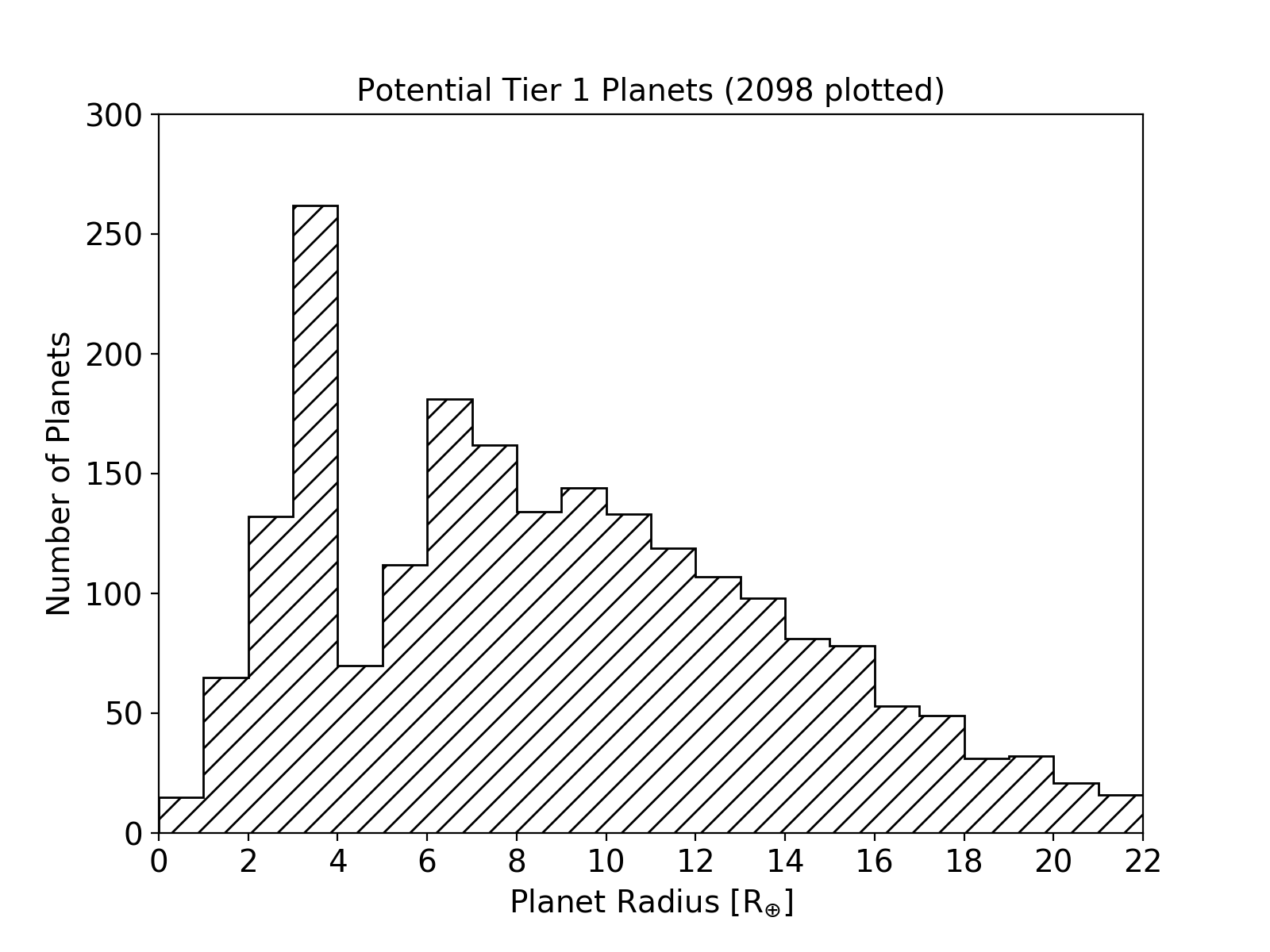}
    \includegraphics[width = 0.37\textwidth]{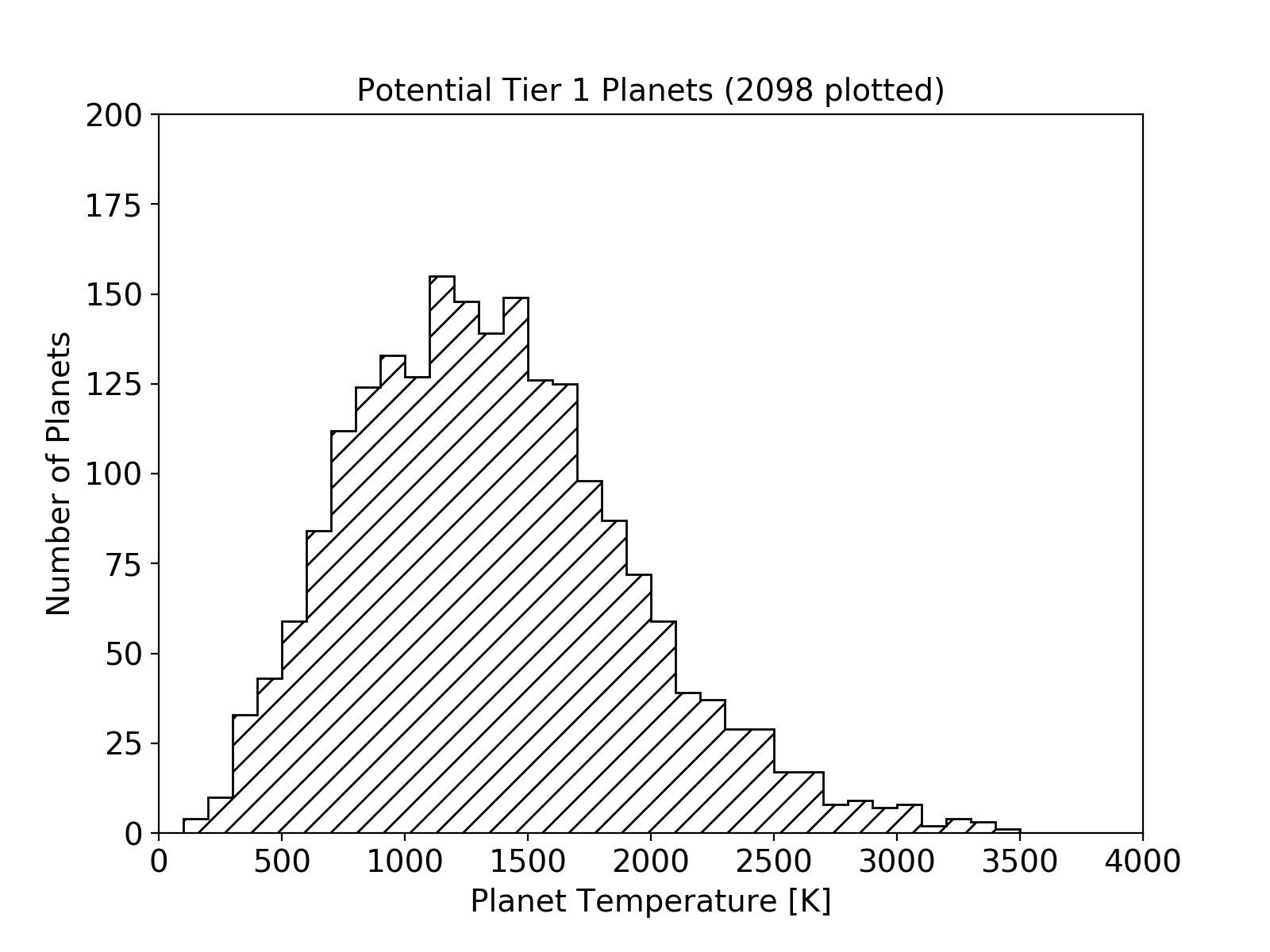}
    \includegraphics[width = 0.37\textwidth]{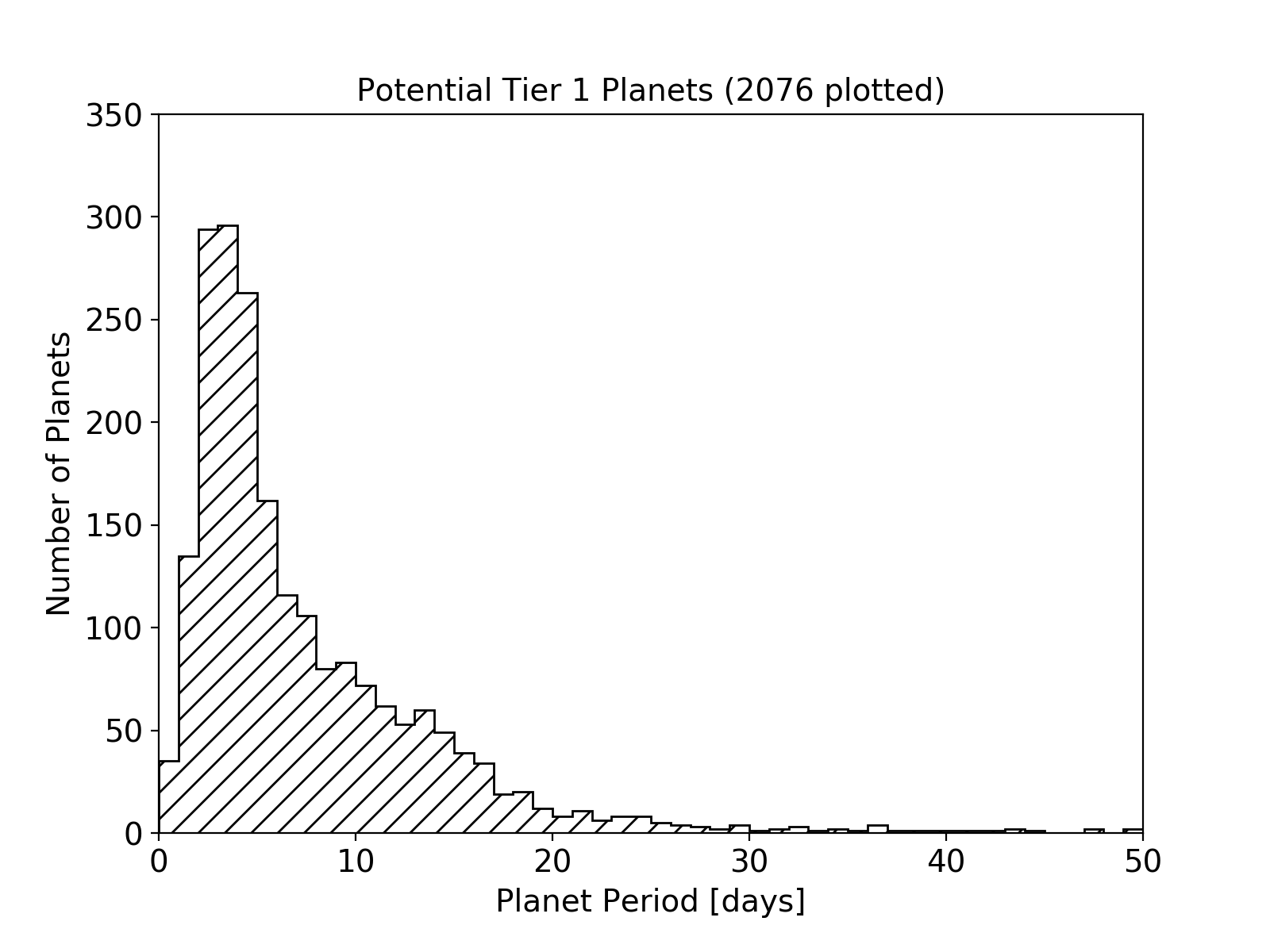}
    \includegraphics[width = 0.37\textwidth]{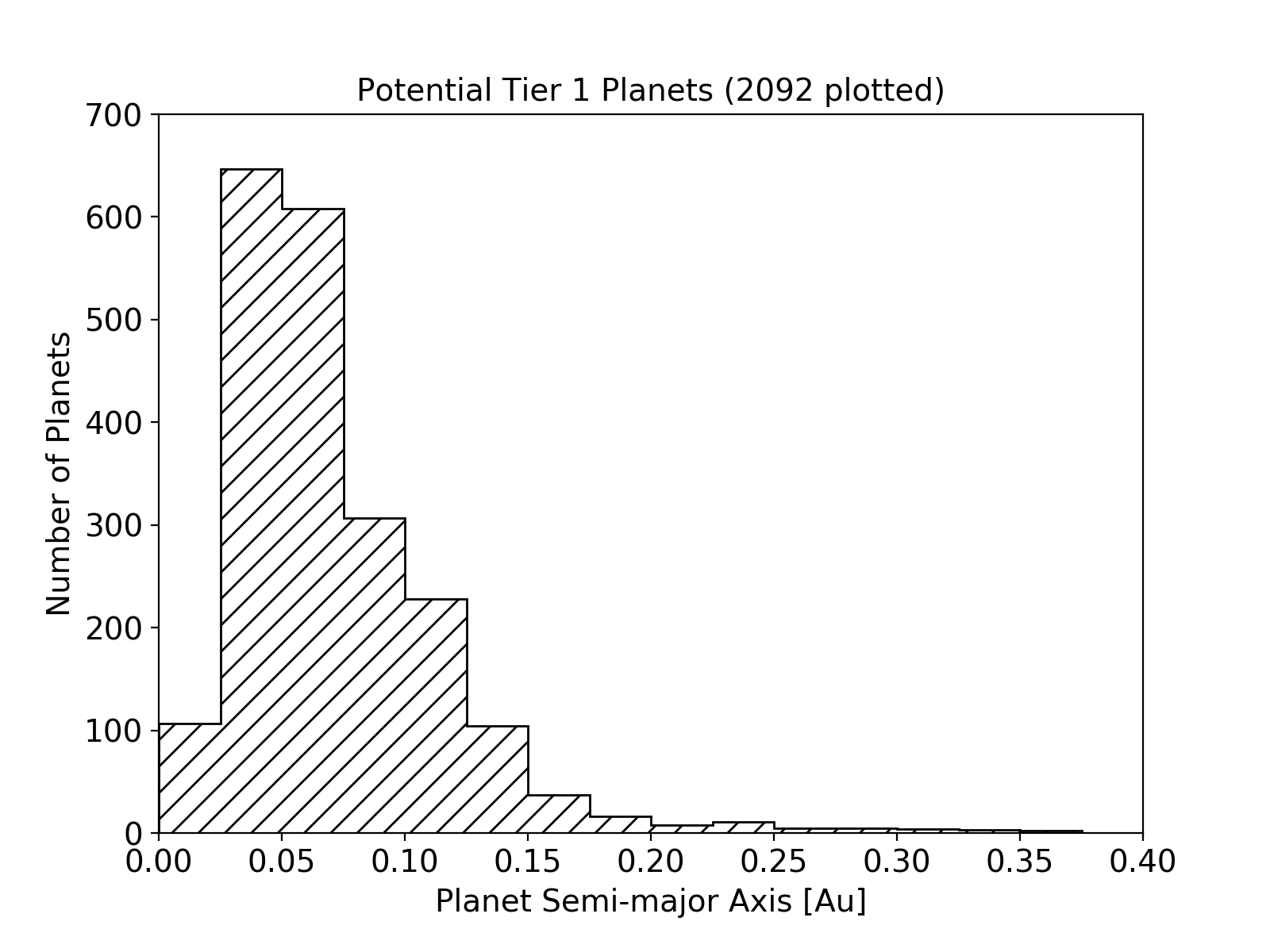}
    \caption{Histograms of the planetary properties within the potential Ariel Tier 1 Catalogue. In some cases, not all planets are plotted for aesthetic reasons.}
    \label{planet para}
\end{figure}

From the noise models created by ArielRad, the catalogue of known and predicted planets was cut down to those for which an SNR $\ge$ 7 could be achieved on the atmosphere within a reasonable number of transits or eclipses. For Tier 1, there are over 2000 potential planets for which the science requirements can be reached with 5 observations or less, far more than the 1000 that will make up the MRS. Being over-saturated in the number of possible targets is useful as it allows for redundancy in the scheduling of observations and it means there is a large catalogue of planets to draw from to allow for a diverse sample to be observed. The distribution of various stellar and planetary parameters for these potential Tier 1 targets is shown in Figures \ref{stellar para} and \ref{planet para}. These show that (i) to achieve a sample of $\sim$1000 planets, Ariel does not need to observe faint stars (except for special targets of interest) (ii) there is a large diversity in planet temperature and radius (iii) the stellar type of planet hosting stars is varied although FG stars are more dominant (iv) the majority of potential targets are located within a few hundred parsecs  (v) most potential targets are close to their stars and have orbits of under 20 days (vi) although the metallicities of many of the host stars is unknown, there is a wide range of values included in the sample.

\begin{figure}[]
  \centering
  \includegraphics[width=0.48\textwidth]{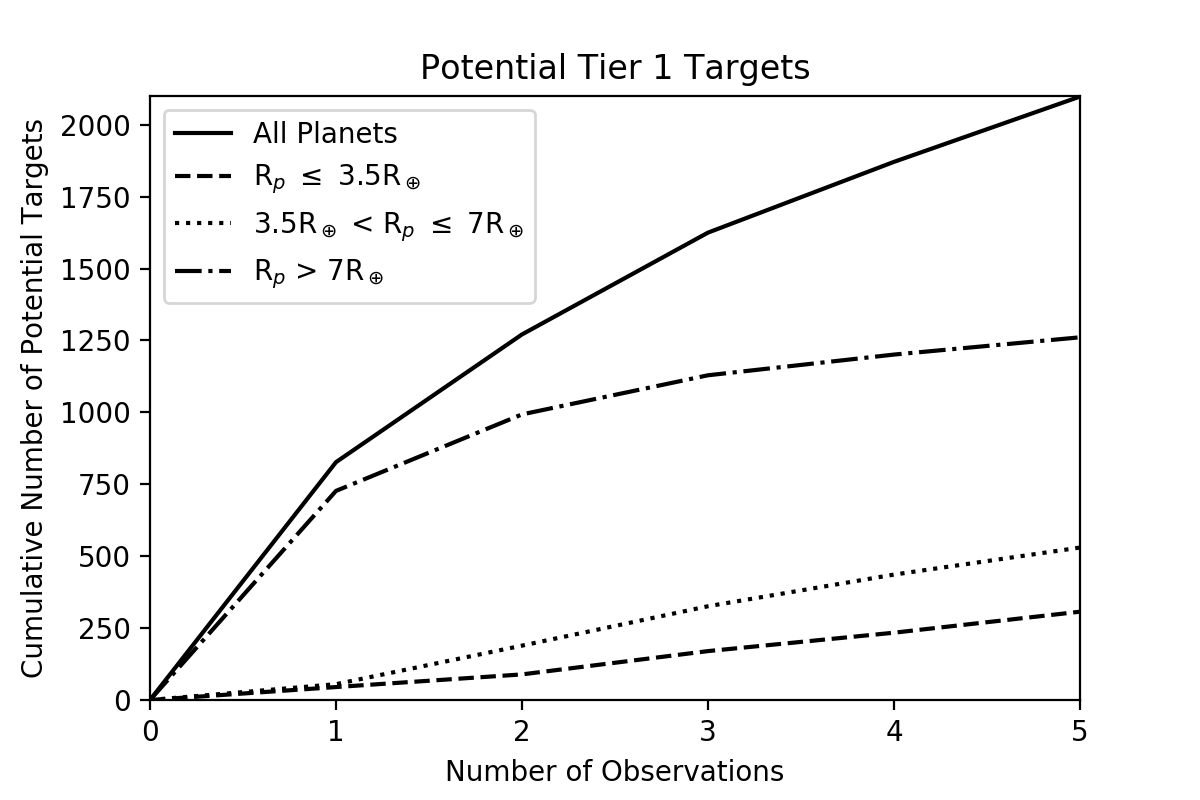}
  \includegraphics[width=0.48\textwidth]{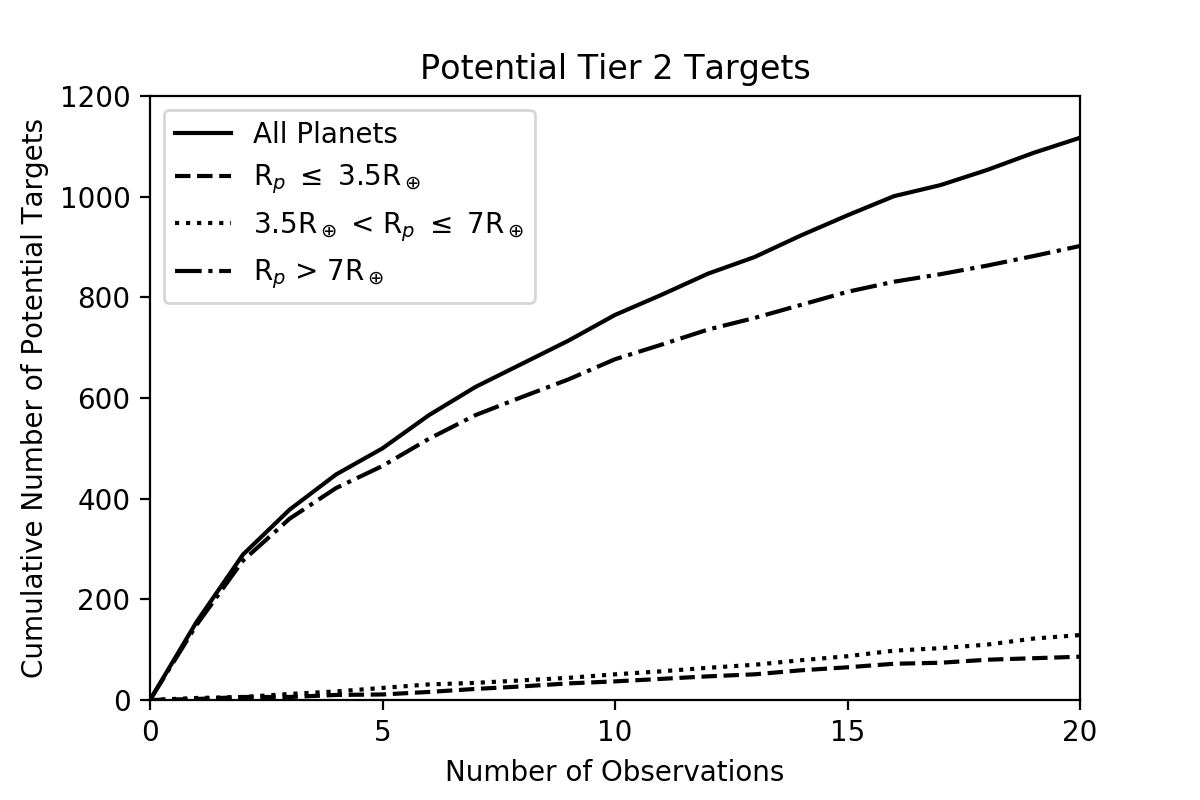}
  \caption{Cumulative number of planets that can be observed in Tiers 1 (left) and 2 (right) with a given number of transits or eclipses.} 
  \label{num obs}
\end{figure}

\begin{figure}[]
  \centering
  \includegraphics[width=1\textwidth]{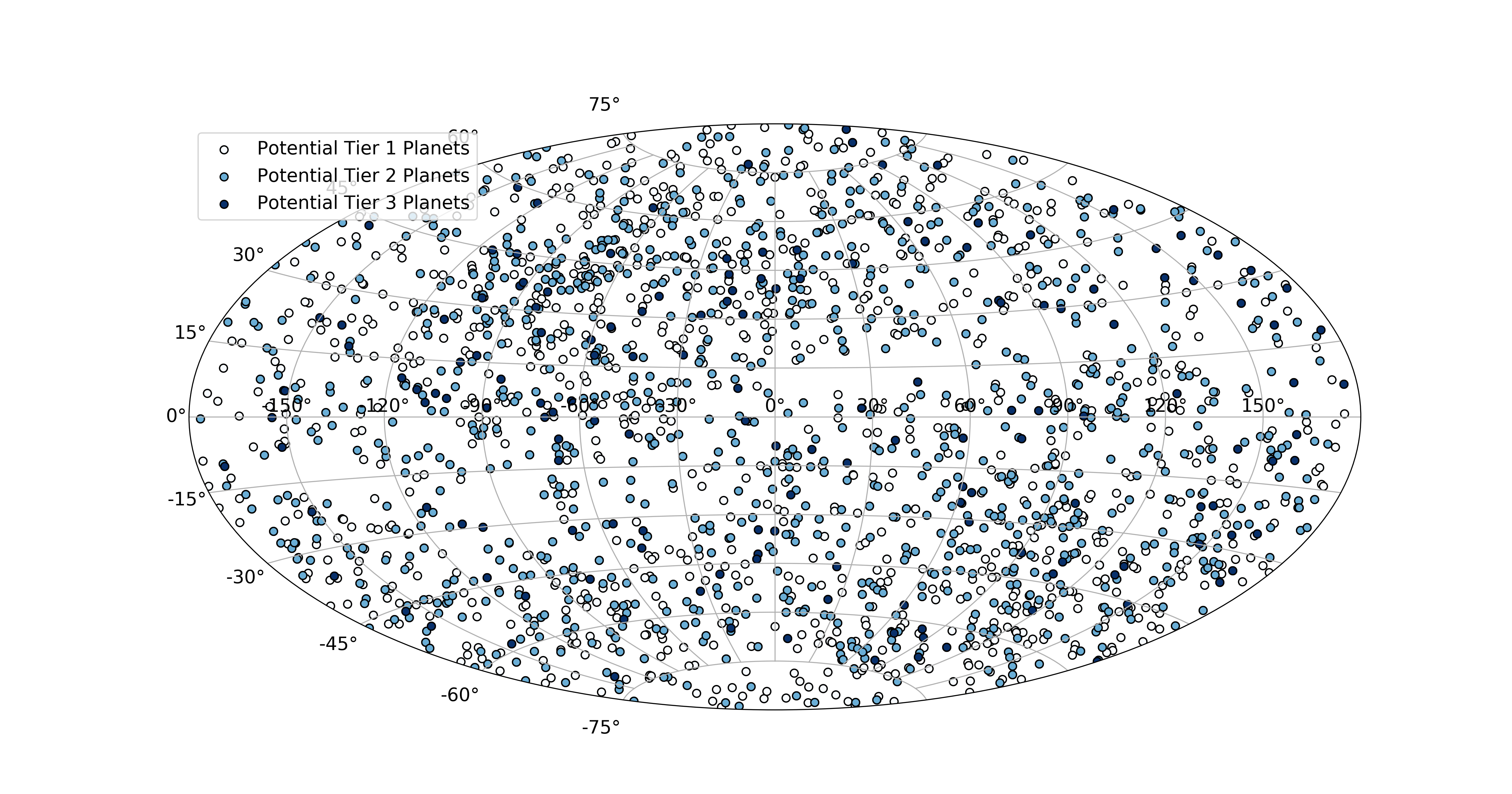}
  \caption{Sky locations of potential targets for study with Ariel. Having targets scattered across the entire sky is beneficial for the scheduling of observations.} 
  \label{sky loc}
\end{figure}

Additionally, $\sim$1000 planets are found to be potentially observable in Tier 2 and Figure \ref{num obs} details the distribution of the number of observations required for these planets as well as those in Tier 1. We find that the number of observable Jupiters (R$_p$ $>$ 7R$_\oplus$) is approaching saturation at 5 observations while the number of suitable smaller planets are rising with increased observations. Ariel will have constant visibility of the ecliptic poles with a partial visibility of the whole sky at lower latitudes. The sky location of possible planets for study in each tier with Ariel is shown in Figure \ref{sky loc} and they are found to be well distributed across the sky but with a noticeable gap close to the ecliptic due to a lack of TESS coverage in its primary mission. A table of the currently-known exoplanets which are suitable for study with Ariel is included in the Appendix.

\clearpage

\subsection{Creation of an Example Mission Reference Sample}
\label{creationmrs}

Ariel has a nominal life of 4 years (extendable to 6) including a 6-month commissioning and calibration phase. Additionally, scheduling constraints, such as telescope housekeeping, slewing between targets and data down-link, reduce the available science time. Ariel will therefore have $\sim$3 years of usable science time during its nominal life. Having established that there will be a large number of planetary atmospheres that are suitable for characterisation with Ariel we explore the number that could be observed over the mission lifetime.

The approach adopted during Phase A consisted of choosing a very diverse, and as complete as possible, combination of star/planet parameters while minimising the number of repeated observations by selecting the planets around the brightest stars. Here, we chose three main parameters to classify the potential targets by: stellar effective temperature, planetary radius and planetary equilibrium temperature. Each parameter is split into a number of classes and Table \ref{tab:classification} summarises these distinctions. We bin the planets by these 3 parameters, and where possible, ensure that at least 2 planets within each bin are contained within the Mission Reference Sample. Future selections will also classify planets by their density and the metallicity of the host star. These five basic characteristics are thought to have a large impact on the chemistry and thus choosing planets with a broad range in these parameters should yield a multifarious exoplanet population for study.

\begin{table}[h]
    \centering
    \begin{tabular}{lll}
    \hline
    Parameter  & Class  & Bounds  \\ \hline
\multirow{4}{*}{Stellar Effective Temperature} 
            & M  & T$_s$ \textless 3955K\\
            & K  & 3955K \textless T$_s$ \textless 5330K \\
            & G  & 5330K\textless T$_s$ \textless 6070K  \\
            & F  & 6070K \textless T$_s$ \textless 7200K \\ \hline
\multirow{5}{*}{Planetary Radius}                 
& Earth/Super-Earth & R$_p$ \textless 1.8 R$_\oplus$ \\
& Sub-Neptune & 1.8R$_\oplus$ \textless R$_p$ \textless 3.5R$_\oplus$\\
& Neptune & 3.5R$_\oplus$ \textless R$_p$ \textless 6R$_\oplus$ \\
& Jupiter & 6R$_\oplus$ \textless R$_p$ \textless 16R$_\oplus$  \\
& Massive Jupiter   & R$_p$ \textgreater 16R$_\oplus$ \\ \hline
\multirow{5}{*}{Planetary Equilibrium Temperature} 
& Temperate/Warm & T$_p$ \textless 500K  \\
& Very Warm & 500K \textless T$_p$ \textless 1000K \\
& Hot & 1000K \textless T$_p$ \textless 1500K    \\
& Very Hot & 1500K \textless T$_p$ \textless 2500K    \\
& Ultra Hot & T$_p$ \textgreater 2500K \\ \hline
    \label{tab:classification}
    \end{tabular}
    \caption{Bounds used to classify potential planets to ensure a varied population of planets within the Mission Reference. Sample}
    \label{tab:classification}
\end{table}

\begin{figure}[h!]
  \centering
  \includegraphics[width=1\textwidth]{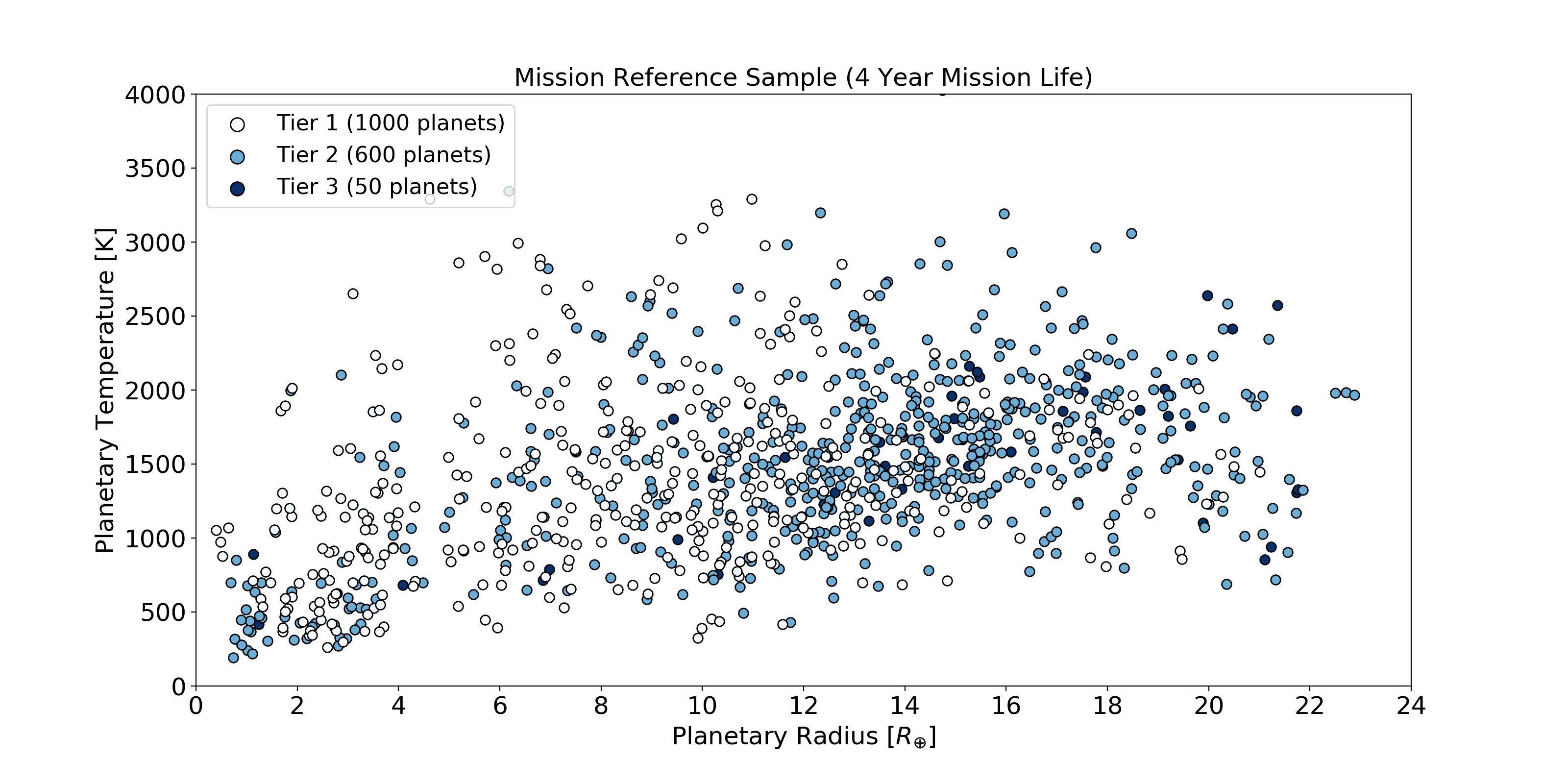}
  \caption{Planetary radius and temperature distribution of a potential Ariel mission reference sample from ArielRad} \label{ArielRad MRS}
\end{figure}

\begin{figure}[h!]
  \centering
  \includegraphics[width=1\textwidth]{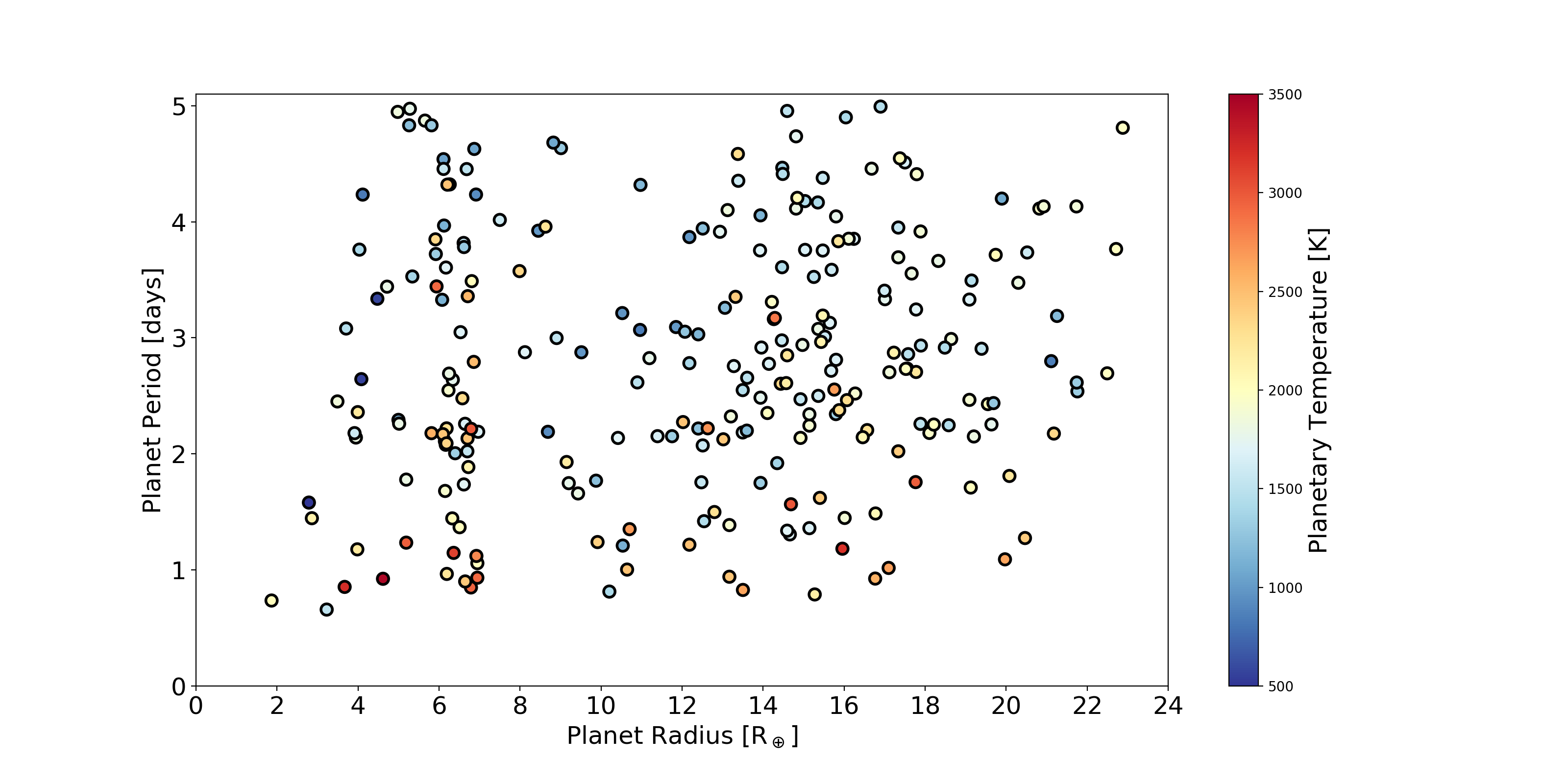}
  \caption{Potential phase-curve targets for Ariel. The colour of points highlights the planetary equilibrium temperature. Spectroscopic phase-curves should be possible for Jupiter-sized planets while smaller planets are suitable for multi-band photometric observations.}
  \label{phasecurves}
\end{figure}

Adopting this strategy we obtain a distribution of planets by radius and temperature as displayed in Figure \ref{ArielRad MRS}. Planets selected for Tier 3 are also included in Tier 2 and, in turn, Tier 1 planets incorporate all those studied in Tier 2. Although not considered in-depth here, 10\% of mission time is reserved for Tier 4 and we highlight potential targets for phase-curves in Figure \ref{phasecurves}. For larger planets, these are those which can easily be observed at Tier 2 resolutions in both transit and eclipse while for smaller planets, it is those that can be studied at Tier 1 resolutions in both methods. Phase-curve targets are also required to be on relatively short orbits and thus are generally found to be hot (or very-hot).

Different observing strategies have been discussed within the Ariel team including acquiring data in both transit and eclipse for some Tier 2 planets. Such observations would increase our ability to characterise the atmospheres of these targets but would reduce the total number of planets studied. Table \ref{tab: mrs times} highlights the science time (i.e. time on target) required to achieve different observations. These discussions are ongoing and further studies will be undertaken but it can be seen that acquiring data in the secondary method (i.e. the method which gives a lower SNR) for some of the best planets will not require significant mission time. However, the total number of Tier 2 planets may have to be sacrificed to achieve this.

\begin{table}[h]
    \centering
    \begin{tabular}{lll}
    \hline
    Number of Planets & Observation Requirement & Required Science Time [hours] \\ \hline
    1000 & Achieve Tier 1 resolutions & $\sim$10,600 \\ \hline
    400 & \multirow{3}{*}{Increase resolution from Tier 1 to Tier 2}  & $\sim$3,100 \\
    500 &  & $\sim$6,000 \\
    600 & & $\sim$10,500 \\ \hline
    200 & \multirow{3}{*}{Achieve Tier 1 resolutions in second method} & $\sim$1,400 \\
    300 &  & $\sim$2,500 \\
    400 &  & $\sim$4,200 \\ \hline
    50 & Tier 3 (5 repeated observations per planet) & $\sim$1,700 \\\hline
    - & Tier 4 (additional science time) & $\sim$2,300 \\\hline
    \end{tabular}%
    \caption{Mission time required to achieve different observation goals. The total science time over the 4 year primary life is $\sim$24,800 hours. Note that for some bright targets (e.g. HD 209458 b), Tier 2 or 3 resolutions would be reached in a single observation.}
    \label{tab: mrs times}
\end{table}

Hence, ArielRad simulations combined with the TESS yield suggested by \cite{barclay} predict that Ariel will be able to observe 1000 planets within the primary mission (e.g. Figure \ref{ArielRad MRS}). The number of planets within this updated version of Mission Reference Sample is similar to that of the Phase A study although we find an increase in the number of Tier 2 planets compared to the results of \cite{zingales} on top of the 10\% mission lifetime dedicated to Tier 4 planets. Therefore, from the input catalogue of currently-known and predicted planets, ArielRad simulations suggest Ariel should be more than capable of achieving the science requirement of characterising the atmospheres of hundreds of diverse extra-solar planets.


\section{Characterisation of Small Planets}

Section \ref{ArielRad} shows that from the catalogue of known planets and predicted TESS detections ArielRad produces a Mission Reference Sample consistent with that created in Phase A with the ESA radiometric model and predicted targets by \cite{zingales}. Choosing the Mission Reference Sample in this way, naturally leads to a proportionally larger number of gaseous planets being selected for observation. However, warm and hot Super-Earths (and Earth-sized planets) are well within Ariel's capabilities, especially given that they are expected to be more numerous around bright stars.

Smaller planets, particularly those which could be rocky, are an intriguing population of bodies, especially since the discovery of the ``Fulton gap'' at $\sim$1.8R$_{\oplus}$ by the California-Kepler Survey (CKS) \citep{fulton}. This distribution seemingly indicates two populations of small planets:  those which have retained a volatile dominated atmosphere and those which are expected to have lost this more primordial envelope (e.g. \cite{owen}) or never had one. Characterising the atmospheres of planets with radii smaller than 3.5 $R_{\oplus}$, and in particular those within the transition region from rocky to gaseous, is fundamental in uncovering the nature of this population and would be very informative for planetary formation and evolution theories. More specifically, understanding whether the atmosphere is still primordial (i.e. H/He rich, possibly thick) or more evolved (i.e. richer in heavier elements, thin or completely absent) may constrain formation (formed in situ or remnants of more massive bodies which have migrated to closer orbits) and evolution scenarios (e.g. hydrogen escaped, a secondary atmosphere which might hint at the interior composition).

Here we explore a different option for the Ariel MRS, with more emphasis on the interpretation of the nature of smaller planets, by specifically devoting mission lifetime to studying this dichotomy of small worlds.

In the Mission Reference Sample studied in Section \ref{ArielRad}, $\sim$110 planets with a radius less than 3.5 Earth radii were selected for study over around 600 observations ($\sim$2100 hours of science time) in all three tiers. These planets are located on both sides of the ``Fulton gap''. A key goal of Tier 1 is to discover the fraction of small planets with hydrogen/helium envelopes. For this reason, the number of required observations to detect an atmosphere is estimated assuming a low mean molecular weight so that if a planetary atmosphere has a primordial composition, this atmosphere should be detected with high confidence. Additionally, the atmospheric trace gases should be accurately constrained if the planet is observed in Tier 2 or 3. If no detection is made, the planet either has (i) an atmosphere with a higher molecular weight or (ii) opaque clouds across all wavelengths or (iii) no atmosphere at all.

In all likelihood, some fraction of these planets will have far heavier atmospheres (higher mean molecular weight) and thus will be harder to characterise, requiring more observations to obtain the observational requirements in each tier. In particular, additionally to the H/He atmospheric content, the fraction of H$_2$O present in an atmosphere is also very important to constrain formation/evolution scenarios and the delivery of volatiles to the inner part of the planetary system. ``Water worlds'', i.e. planets with a significant amount of H$_2$O on their surface or in the subsurface (e.g. \cite{leger}), or magma ocean planets with a steam atmosphere (e.g. \cite{hamano}), are expected to have atmospheres with a large fraction of H$_2$O.

However, the characteristics of a planet's atmosphere (if present) cannot be known before observations are undertaken, unless these targets are observed previously with other facilities from space or the ground. To quantify the fraction of lifetime needed to characterise the atmospheric composition of small planets with an atmosphere heavier than H/He, we select the small planets (R$_p$ $<$ 3.5$R_{\oplus}$) from the example MRS for further study. The science time required to achieve Tier 1 resolutions (with SNR $>$ 7) for different atmospheric compositions is determined and compared to the Tier 1 time assumed in Section \ref{MRS} (Table \ref{tab: small planets}).

\begin{table}[h]
    \centering
    \begin{tabular}{lll}
    \hline
    Atmospheric Mean Molecular Weight & Number of Planets & Required Science Time [hours] \\ \hline
    2.3 & All & $\sim$1,000 (t$_0$) \\ \hline
    5 & 50 & t$_0$ + $\sim$360 \\
     & All & t$_0$ + $\sim$3000 \\ \hline
    8 & 50 & t$_0$ + $\sim$1,100 \\
     & All & t$_0$ + $\sim$9,200 \\ \hline
    10 & 50 & t$_0$ + $\sim$1,900 \\ \hline
    15 & 50 & t$_0$ + $\sim$4,400 \\ \hline
    18 & 25 & t$_0$ + $\sim$1,700 \\
     & 50 & t$_0$ + $\sim$6,400 \\ \hline
    28 & 25 & t$_0$ + $\sim$4,300 \\
     & 50 & t$_0$ + $\sim$15,600 \\ \hline
    \end{tabular}%
    \caption{Mission time required to achieve Tier 1 resolutions (at SNR \textgreater 7) for the 113 small planets in the example MRS assuming different mean molecular weights. The total science time over the 4 year primary life is $\sim$24,800 hours. t$_0$ is the time spent observing small planets in Tier 1 of the standard MRS.}
    \label{tab: small planets}
\end{table}

\begin{figure}
    \centering
    \includegraphics[width = 0.85\textwidth]{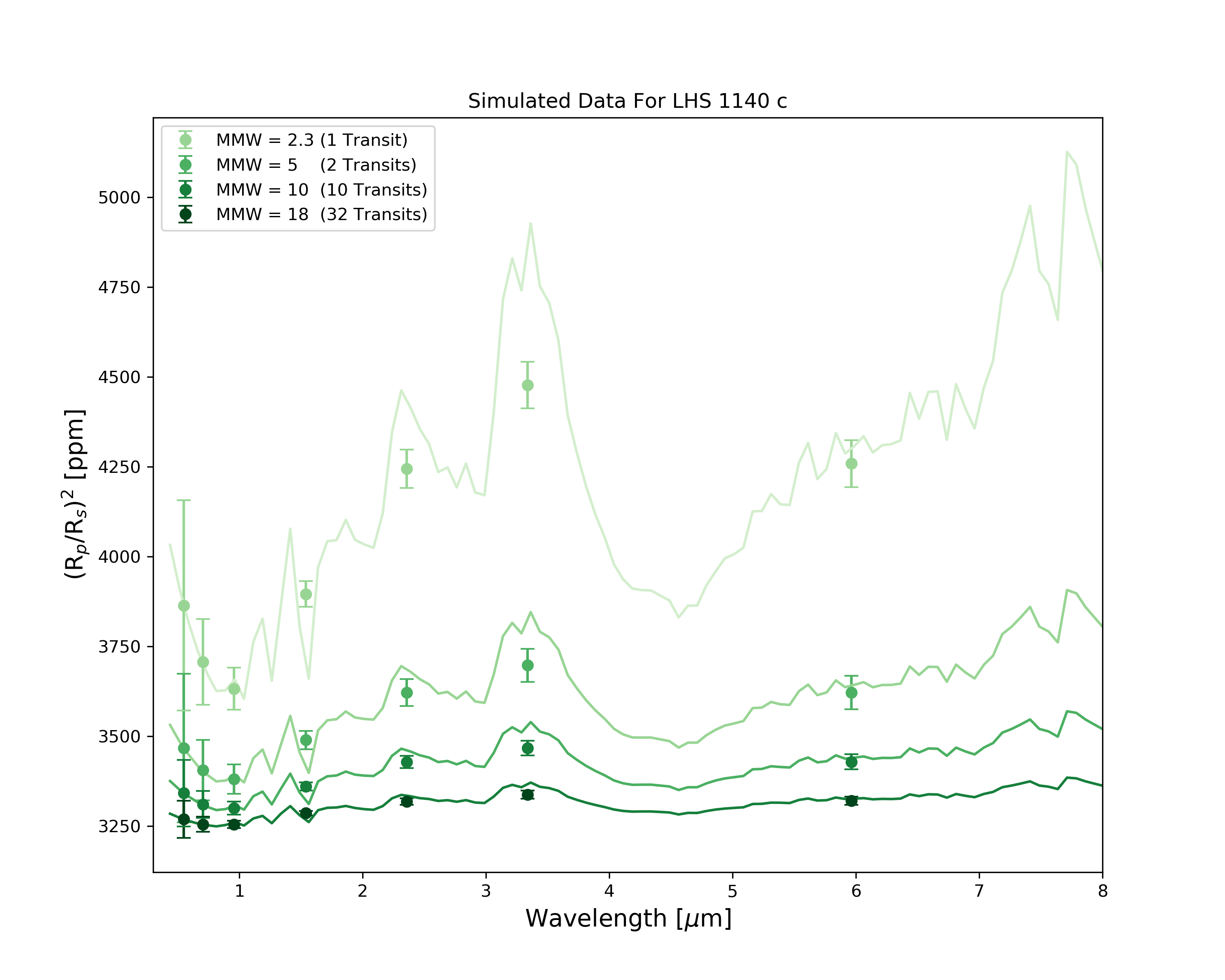}
    \caption{Simulated Tier 1 data of LHS 1140 c for different atmospheric weights. The atmosphere is modelled with 10$^{-5}$ of H$_2$O and CH$_4$ and the mean molecular weight is varied by modifying the nitrogen ratio. The number of transits quoted is the requirement for an SNR \textgreater 7 to be achieved on the atmosphere at Tier 1 resolutions.}
    \label{LHS spectra}
\end{figure}

As expected, the required number of observations (and thus science time) rises with the increasing atmospheric weight. While the atmospheres of smaller planets will be easily probed if H/He dominated, heavier atmospheres would require significant mission time to observe. Distinguishing between primary and secondary atmospheres should be possible for all small planets studied here within a reasonable science time. However, the assumed noise floor of 20ppm limits the characterisation at very high mean molecular weights where the signals become increasingly small. Smaller, cooler planets may also have a nitrogen based atmosphere and we find that, for the Earth-sized planets below 500K in this chosen sample, 25-130 transits would be required to achieve Tier 1 resolutions if the atmospheres had a molecular weight of 28. Figure \ref{LHS spectra} shows simulated data for one such planet, LHS 1140 c \citep{ment}, for various atmospheric weights. The dampening in the spectra due to a heavier atmosphere can clearly be seen. Generally, the best targets could be easily characterised regardless of their atmospheric composition while for others achieving the required signal uncertainty will be difficult if the atmosphere is dense. We note that the impact of clouds is expected to be well captured in the simulations for higher mean molecular weight, where signals are up to fourteen times smaller than the ones for atmospheres which are cloud-free and H/He-rich. Additional observations of the planet at different phases may provide further constraints on the cloud types and distribution (e.g. \cite{charnay}). Observations of smaller planets could be undertaken in a tiering style system where the data is analysed after several visits with decisions made on continuing the observations based on the results seen. Science goals for such an observing strategy could include the determination of whether an atmosphere is primary, secondary or not present.

From this preliminary study, we appreciate that providing significant time to observe smaller planets would be valuable for their more in-depth chemical/cloud characterisation after an initial survey. Here we have presented a possible option including the in-depth analysis of $\sim$110 small planets, but of course different combinations of strategies could and will be considered in this and future mission Phases to optimise the breadth and depth of the Ariel sample during its mission lifetime and prioritise its science objectives. If much of the primary mission is dedicated to an in-depth survey of smaller planets, the total number of planets observed by Ariel would be reduced. Hence, some of the more speculative questions could be left for a potential extended mission. This study shows that Ariel has the potential to characterise the atmospheres of planets of all sizes. Data from such a multifarious population would be invaluable for our knowledge of planetary formation and evolution.

\section{Discussion}

\subsection{Dependence of Predicted Yields on the Accuracy of Planetary Occurrence Statistics}

Here, expected TESS detections have been used to estimate the number, and type, of exoplanets that could be potential targets for ARIEL. Predicted yields for future missions are, of course, speculative in natural and highly dependent on the assumptions of the study. In \cite{barclay} the planetary occurrence statistics for AFGK stars were taken from \cite{fressin} and from \cite{dressing} for M dwarfs. More recent studies may suggest higher occurrence rates for some classes of small planets (e.g. \cite{mulders,fulton}) and the differences between these could affect the yield of TESS. The NASA Exoplanet catalogue currently contains 9 confirmed TESS planets and many more are yet to be added. Comparing these first detections to the expected yield does not result in any large discrepancies and all these current TESS planets are found to be excellent targets for study with Ariel (see Table \ref{TESS}). However, with only a handful of confirmed detections it is too early to speculate on the accuracy of the predicted yield. Further constraints on the occurrence of planets on short periods is likely to be a key outcome of the TESS mission, particularly for M dwarfs. In any case, the primary mission of TESS is due to finish in 2020 and thus the yield from this mission will be known long before Ariel launches.

\subsection{Scheduling of Observations}

Here, the scheduling of observations has not been considered although studies in the Ariel consortium are being undertaken which will utilise the Mission Reference Sample \citep{garcia,morales}. Such studies will provide a greater understanding of the impact of scheduling constraints (telescope housekeeping, slewing between targets etc.) and a key issue may be observation overlaps (i.e. two planets transiting at the same time). Having additional, back-up targets is likely to be useful for scheduling purposes and this study shows that there should be an over-saturation of suitable planets for characterisation. The list of potential targets constructed here will be used as an input for such efforts.

\subsection{Tiering System for Smaller Planets}

The ambiguity in the atmospheric composition of smaller planets results causes complexities when planning via the originally proposed three tier observing structure. Additionally, the major constituents of an atmosphere could be recovered at resolutions below that of Tier 2. Therefore, a separate tiering system for smaller planets which is based around confirming the presence (or absence) of a clear atmosphere of a given mean molecular weight may be required. Once the catalogue of potential planets is completely formed of known planets, additional considerations in the selection of smaller planets such as photo-evaporation and isolation flux (e.g. \cite{owen,swain}) will need to be taken into account to ensure a diverse population of planets are studied.

\subsection{Next Steps and Final Selection of the Mission Reference Sample}

The Mission Reference Sample presented here is merely one example of a population of planets that Ariel could observe. The selection of the final list of targets will require far more discussion and input from scientists from across the exoplanet community, particularly as the predicted planets from this list are replaced with actual detections. In the coming years, observations with current ground and space-based facilities (e.g. VLT, \textit{Hubble}, \textit{Spitzer}) and future observatories (e.g. E-ELT \citep{brandl}, JWST \citep{greene}, Twinkle \citep{edwards_exo}) will further characterise the atmospheres of the known exoplanet population. These studies will increase our knowledge of these distant worlds and may begin to highlight trends in atmospheric chemistry. Such insights will inevitably be used to optimise the Ariel Mission Reference Sample, maximising the synergies between different facilities, and to this end a website has been created to host the list of potentially observable planets \footnote{\url{https://arielmission.space/target-list/}}. This open-access site will contain all available datasets on these planets, highlighting planetary systems for which further characterisation would be beneficial (e.g. refinement of ephemerides or stellar parameters) and providing the chance for the entire community to contribute to the Ariel target selection. Therefore, Ariel will embrace the exoplanet community by offering open involvement in the observation planning process as well as providing regular timely public releases of high quality data products at various processing levels throughout the mission. Additionally, targets within the list are being used as the basis for simulated data in several data challenges organised to engage the exoplanet community in Ariel\footnote{\url{https://ariel-datachallenge.azurewebsites.net}}. These efforts, particularly the continuous dialogue with the wider community, will ensure that the Ariel observation strategy facilitates the maximum possible science yield for the entire exoplanet field.

\section{Conclusions}

An updated analysis of the currently-known planets and predicted TESS discoveries, as well as Ariel predicted performances, supports and improves the conclusions of the previous Mission Reference Sample (MRS) study from Phase A: Ariel will be capable of characterising 1000 exoplanet atmospheres during the primary mission life. The total number of potential planets to choose this MRS from is found to be over 2000 meaning there is a surplus of targets. Within this list of planets there is a large range of planetary and stellar parameters, ensuring that the MRS is diverse; a key requirement for meeting Ariel's mission objectives. The example MRS selected here allows for 1000 planets to be studied, with high-quality spectroscopic data being obtained for 600 of these during the 4 year primary mission life. The selection also reserves mission time for other observation strategies (Tier 4). These could include phase-curves, non-transiting planets or targets of interest which are not captured by the current tier system.

Additionally we have explored the mission capability to perform an in depth analysis of small planets' atmospheres, which are expected to be more diverse compared to the gaseous ones. Given the increased observational difficulty to probe atmospheres heavier than H/He, significant mission time may have to be allocated to this task. Trade-offs between studying more planets, observing fewer targets but in greater detail, and/or choosing interesting planets which require more observational time, will form a key part in the selection of the final Mission Reference Sample. Generating an optimal catalogue of potential candidates is key in these efforts and this list of targets will be constantly updated with new planet discoveries.

\vspace{5mm}
\noindent\textbf{Acknowledgements}

The authors wish to thank the members of the European Space Agency Ariel SAT and the Ariel consortium for useful comments and suggestions. This research has made use of the NASA Exoplanet Archive, which is operated by the California Institute of Technology, under contract with the National Aeronautics and Space Administration under the Exoplanet Exploration Program. Additionally, the Open Exoplanet Catalogue, TEPCat and exoplanet.eu have been utilised as supplementary data sources. This work has been funded through the ERC Consolidator grant ExoLights (GA 617119) and the STFC grants ST/P000282/1, ST/P002153/1 and ST/S002634/1.

\section{Appendix}
\centering

\begin{longtable}{lllllll}

\caption{Currently-known exoplanets which are considered here to be potential targets for Ariel. This list will continue to evolve as surveys discover more planets.}\\ \hline
  & \multicolumn{2}{l}{Star Properties} & \multicolumn{2}{l}{Planet Properties} & Maximum & Preferred \\
 Planet Name & Eff Temp [K] & K Magnitude &  Radius [R$_\oplus$] & Equil Temp [K] & Tier & Method \\ \hline
\endfirsthead

\multicolumn{5}{c}%
        {{\bfseries Table \thetable\ Continued from previous page}} \\
        \hline
        
& \multicolumn{2}{l}{Star Properties} & \multicolumn{2}{l}{Planet Properties} & Maximum & Preferred \\
 Planet Name & Eff Temp [K] & K Magnitude &  Radius [R$_\oplus$] & Equil Temp [K] & Tier & Method \\ \hline
 \endhead
 
 \hline
 \endfoot
 
55Cnce	&	5196	&	4.01	&	1.87	&	1997	&	2	&	Transit	\\
CoRoT-10b	&	5075	&	11.78	&	10.64	&	685	&	1	&	Transit	\\
CoRoT-11b	&	6440	&	11.25	&	15.69	&	1771	&	2	&	Eclipse	\\
CoRoT-19b	&	6090	&	11.84	&	14.16	&	1697	&	2	&	Eclipse	\\
CoRoT-2b	&	5625	&	10.31	&	16.09	&	1582	&	3	&	Eclipse	\\
CoRoT-23b	&	5900	&	12.36	&	11.52	&	1679	&	2	&	Eclipse	\\
CoRoT-28b	&	5150	&	11.03	&	10.48	&	1395	&	1	&	Eclipse	\\
CoRoT-3b	&	6740	&	11.62	&	11.08	&	1737	&	1	&	Eclipse	\\
EPIC211945201b	&	6069	&	8.84	&	5.64	&	913	&	1	&	Transit	\\
EPIC246851721b	&	6202	&	9.89	&	11.53	&	1434	&	2	&	Transit	\\
GJ1132b	&	3270	&	8.32	&	1.16	&	636	&	2	&	Transit	\\
GJ1214b	&	3026	&	8.78	&	2.79	&	627	&	3	&	Transit	\\
GJ3470b	&	3600	&	7.99	&	4.48	&	698	&	2	&	Transit	\\
GJ436b	&	3475	&	6.07	&	4.08	&	681	&	3	&	Transit	\\
GJ9827b	&	4269	&	7.19	&	1.59	&	1198	&	1	&	Transit	\\
GJ9827d	&	4269	&	7.19	&	2.03	&	695	&	1	&	Transit	\\
HAT-P-1b	&	5980	&	8.86	&	14.47	&	1353	&	3	&	Eclipse	\\
HAT-P-11b	&	4780	&	7.01	&	4.27	&	848	&	2	&	Transit	\\
HAT-P-12b	&	4650	&	10.11	&	10.52	&	979	&	2	&	Transit	\\
HAT-P-13b	&	5653	&	8.98	&	13.96	&	1685	&	2	&	Eclipse	\\
HAT-P-14b	&	6600	&	8.85	&	15.58	&	1601	&	2	&	Eclipse	\\
HAT-P-15b	&	5568	&	9.64	&	11.63	&	917	&	2	&	Eclipse	\\
HAT-P-16b	&	6158	&	9.55	&	14.14	&	1664	&	3	&	Eclipse	\\
HAT-P-17b	&	5246	&	8.54	&	11.52	&	793	&	2	&	Transit	\\
HAT-P-18b	&	4803	&	10.23	&	10.92	&	867	&	2	&	Transit	\\
HAT-P-19b	&	4990	&	10.55	&	12.42	&	1032	&	2	&	Transit	\\
HAT-P-20b	&	4595	&	8.6	&	9.51	&	990	&	3	&	Transit	\\
HAT-P-21b	&	5588	&	10.11	&	12.18	&	1303	&	2	&	Transit	\\
HAT-P-22b	&	5302	&	7.84	&	12.62	&	1307	&	3	&	Eclipse	\\
HAT-P-23b	&	5905	&	10.79	&	11.96	&	2094	&	2	&	Eclipse	\\
HAT-P-24b	&	6373	&	10.54	&	14.27	&	1672	&	2	&	Eclipse	\\
HAT-P-25b	&	5519	&	10.82	&	12.45	&	1209	&	2	&	Eclipse	\\
HAT-P-26b	&	5079	&	9.58	&	6.91	&	1013	&	2	&	Transit	\\
HAT-P-27b	&	5300	&	10.11	&	11.19	&	1236	&	2	&	Eclipse	\\
HAT-P-28b	&	5680	&	11.1	&	13.3	&	1412	&	2	&	Eclipse	\\
HAT-P-29b	&	6087	&	10.3	&	12.84	&	1285	&	2	&	Eclipse	\\
HAT-P-3b	&	5185	&	9.45	&	10.31	&	1184	&	2	&	Eclipse	\\
HAT-P-30b	&	6304	&	9.15	&	15.8	&	1675	&	3	&	Eclipse	\\
HAT-P-31b	&	6065	&	10.08	&	11.96	&	1408	&	2	&	Transit	\\
HAT-P-32b	&	6207	&	9.99	&	19.2	&	1823	&	3	&	Eclipse	\\
HAT-P-33b	&	6446	&	10	&	20.3	&	1816	&	3	&	Eclipse	\\
HAT-P-34b	&	6442	&	9.25	&	14.81	&	1333	&	2	&	Eclipse	\\
HAT-P-36b	&	5560	&	10.6	&	12.62	&	1858	&	2	&	Eclipse	\\
HAT-P-37b	&	5500	&	11.67	&	12.93	&	1306	&	2	&	Eclipse	\\
HAT-P-38b	&	5330	&	10.5	&	9.05	&	1101	&	1	&	Transit	\\
HAT-P-39b	&	6430	&	11.16	&	17.24	&	1789	&	2	&	Eclipse	\\
HAT-P-4b	&	5860	&	9.77	&	13.98	&	1472	&	2	&	Eclipse	\\
HAT-P-40b	&	6080	&	10.01	&	16.68	&	1805	&	2	&	Eclipse	\\
HAT-P-41b	&	6390	&	9.73	&	22.5	&	1981	&	3	&	Eclipse	\\
HAT-P-42b	&	5743	&	10.63	&	14.05	&	1460	&	2	&	Eclipse	\\
HAT-P-43b	&	5645	&	11.76	&	14.06	&	1386	&	2	&	Eclipse	\\
HAT-P-44b	&	5295	&	11.28	&	13.63	&	1130	&	2	&	Transit	\\
HAT-P-45b	&	6330	&	10.2	&	15.65	&	1686	&	2	&	Eclipse	\\
HAT-P-46b	&	6120	&	9.92	&	14.09	&	1488	&	2	&	Eclipse	\\
HAT-P-49b	&	6820	&	9.35	&	17.45	&	2171	&	3	&	Eclipse	\\
HAT-P-5b	&	5960	&	10.48	&	13.28	&	1569	&	2	&	Eclipse	\\
HAT-P-50b	&	6280	&	10.5	&	14.13	&	1897	&	2	&	Eclipse	\\
HAT-P-51b	&	5449	&	11.61	&	14.19	&	1216	&	2	&	Transit	\\
HAT-P-52b	&	5131	&	11.62	&	11.07	&	1241	&	2	&	Eclipse	\\
HAT-P-53b	&	5956	&	12.1	&	14.46	&	1818	&	2	&	Eclipse	\\
HAT-P-54b	&	4390	&	10.33	&	10.36	&	839	&	2	&	Eclipse	\\
HAT-P-55b	&	5808	&	11.63	&	12.97	&	1342	&	2	&	Eclipse	\\
HAT-P-56b	&	6566	&	9.83	&	16.57	&	1880	&	2	&	Eclipse	\\
HAT-P-57b	&	6330	&	9.43	&	19.09	&	1895	&	3	&	Eclipse	\\
HAT-P-6b	&	6570	&	9.31	&	16.24	&	1706	&	3	&	Eclipse	\\
HAT-P-65b	&	5835	&	11.53	&	20.74	&	1974	&	2	&	Eclipse	\\
HAT-P-66b	&	6002	&	11.68	&	17.45	&	1944	&	2	&	Eclipse	\\
HAT-P-67b	&	6406	&	8.9	&	22.88	&	1967	&	3	&	Transit	\\
HAT-P-7b	&	6389	&	9.33	&	16.57	&	2272	&	3	&	Eclipse	\\
HAT-P-8b	&	6200	&	8.95	&	15.36	&	1809	&	3	&	Eclipse	\\
HAT-P-9b	&	6350	&	11.02	&	15.36	&	1565	&	2	&	Eclipse	\\
HATS-1b	&	5870	&	10.58	&	14.29	&	1399	&	2	&	Eclipse	\\
HATS-10b	&	5880	&	11.51	&	10.63	&	1435	&	1	&	Eclipse	\\
HATS-11b	&	6563	&	12.24	&	17.66	&	1783	&	2	&	Eclipse	\\
HATS-12b	&	6357	&	11.39	&	7.59	&	1614	&	1	&	Transit	\\
HATS-13b	&	5523	&	11.98	&	13.3	&	1276	&	2	&	Eclipse	\\
HATS-17b	&	5846	&	10.7	&	8.53	&	832	&	1	&	Transit	\\
HATS-2b	&	5227	&	11.39	&	12.82	&	1614	&	2	&	Eclipse	\\
HATS-22b	&	4803	&	10.94	&	10.46	&	877	&	2	&	Transit	\\
HATS-24b	&	6346	&	11.38	&	16.32	&	2121	&	2	&	Eclipse	\\
HATS-25b	&	5715	&	11.42	&	13.83	&	1307	&	2	&	Eclipse	\\
HATS-26b	&	6071	&	11.44	&	19.2	&	1964	&	2	&	Eclipse	\\
HATS-27b	&	6438	&	11.55	&	16.46	&	1693	&	2	&	Eclipse	\\
HATS-29b	&	5670	&	10.88	&	13.73	&	1236	&	2	&	Eclipse	\\
HATS-3b	&	6351	&	10.69	&	15.15	&	1682	&	2	&	Eclipse	\\
HATS-30b	&	5943	&	10.79	&	12.89	&	1446	&	2	&	Eclipse	\\
HATS-31b	&	6050	&	11.57	&	18	&	1867	&	2	&	Eclipse	\\
HATS-33b	&	5659	&	10.29	&	13.5	&	1460	&	2	&	Eclipse	\\
HATS-35b	&	6300	&	11.12	&	16.06	&	2076	&	2	&	Eclipse	\\
HATS-39b	&	6572	&	11.52	&	17.23	&	1683	&	2	&	Eclipse	\\
HATS-4b	&	5403	&	11.61	&	11.19	&	1350	&	2	&	Eclipse	\\
HATS-40b	&	6460	&	12.15	&	17.34	&	2142	&	2	&	Eclipse	\\
HATS-41b	&	6424	&	11.5	&	14.59	&	1716	&	1	&	Eclipse	\\
HATS-43b	&	5099	&	11.56	&	12.95	&	1017	&	2	&	Transit	\\
HATS-45b	&	6450	&	12.14	&	14.11	&	1551	&	1	&	Eclipse	\\
HATS-46b	&	5495	&	11.96	&	9.91	&	1078	&	2	&	Transit	\\
HATS-5b	&	5304	&	10.7	&	10.01	&	1047	&	2	&	Transit	\\
HATS-51b	&	5758	&	10.87	&	15.47	&	1581	&	2	&	Eclipse	\\
HATS-52b	&	6010	&	12.11	&	15.17	&	1922	&	2	&	Eclipse	\\
HATS-6b	&	3724	&	11.22	&	10.95	&	729	&	2	&	Transit	\\
HATS-60b	&	5688	&	10.99	&	12.65	&	1561	&	2	&	Eclipse	\\
HATS-61b	&	5542	&	11.48	&	13.11	&	1252	&	1	&	Eclipse	\\
HATS-64b	&	6554	&	11.7	&	18.42	&	1833	&	2	&	Eclipse	\\
HATS-65b	&	6277	&	11.1	&	16.47	&	1670	&	2	&	Eclipse	\\
HATS-67b	&	6594	&	12.33	&	18.49	&	2240	&	2	&	Eclipse	\\
HATS-68b	&	6147	&	10.95	&	13.52	&	1781	&	2	&	Eclipse	\\
HATS-7b	&	4985	&	10.98	&	6.18	&	1104	&	1	&	Transit	\\
HATS-9b	&	5599	&	11.48	&	13.56	&	1953	&	2	&	Eclipse	\\
HD106315c	&	6277	&	7.85	&	4.22	&	898	&	2	&	Transit	\\
HD149026b	&	6179	&	6.82	&	8.12	&	1715	&	2	&	Eclipse	\\
HD17156b	&	6040	&	6.76	&	12.07	&	904	&	2	&	Transit	\\
HD189733b	&	5052	&	5.54	&	12.4	&	1230	&	3	&	Transit	\\
HD209458b	&	6091	&	6.31	&	15.25	&	1487	&	3	&	Transit	\\
HD219134b	&	4699	&	3.26	&	1.57	&	1040	&	2	&	Transit	\\
HD3167b	&	5528	&	7.07	&	1.67	&	1861	&	1	&	Transit	\\
HD3167c	&	5528	&	7.07	&	2.8	&	630	&	2	&	Transit	\\
HD80606b	&	5561	&	7.32	&	11.74	&	432	&	2	&	Transit	\\
HD89345b	&	5576	&	7.72	&	7.24	&	1114	&	2	&	Transit	\\
HD97658b	&	5175	&	5.73	&	2.3	&	756	&	2	&	Transit	\\
HIP41378e	&	6199	&	7.72	&	5.4	&	530	&	2	&	Transit	\\
HIP41378f	&	6199	&	7.72	&	9.99	&	392	&	2	&	Transit	\\
K2-100b	&	6168	&	9.18	&	3.51	&	1954	&	1	&	Transit	\\
K2-107b	&	6061	&	11.21	&	15.65	&	1845	&	2	&	Eclipse	\\
K2-113b	&	5660	&	11.95	&	11.88	&	1227	&	1	&	Transit	\\
K2-115b	&	5657	&	11.72	&	11.81	&	692	&	1	&	Transit	\\
K2-121b	&	4551	&	10.62	&	7.34	&	807	&	2	&	Transit	\\
K2-129b	&	3459	&	8.85	&	1.02	&	457	&	1	&	Transit	\\
K2-132b	&	4840	&	9.54	&	14.27	&	1578	&	1	&	Eclipse	\\
K2-136c	&	4499	&	8.37	&	2.85	&	558	&	1	&	Transit	\\
K2-139b	&	5370	&	9.66	&	8.92	&	624	&	2	&	Transit	\\
K2-140b	&	5705	&	11	&	11.99	&	1068	&	2	&	Transit	\\
K2-141c	&	4599	&	8.4	&	6.85	&	715	&	3	&	Transit	\\
K2-155c	&	4258	&	9.5	&	2.55	&	544	&	1	&	Transit	\\
K2-18b	&	3457	&	8.9	&	2.33	&	307	&	1	&	Transit	\\
K2-19b	&	5430	&	11.16	&	7.58	&	901	&	1	&	Transit	\\
K2-198b	&	5262	&	9.23	&	4.03	&	698	&	1	&	Transit	\\
K2-199c	&	4648	&	9.62	&	2.72	&	731	&	1	&	Transit	\\
K2-232b	&	6154	&	8.43	&	10.97	&	1015	&	2	&	Transit	\\
K2-233d	&	4950	&	8.38	&	2.59	&	572	&	1	&	Transit	\\
K2-237b	&	6257	&	10.22	&	18.11	&	1952	&	3	&	Eclipse	\\
K2-238b	&	5630	&	12.03	&	14.27	&	1649	&	2	&	Eclipse	\\
K2-24b	&	5625	&	9.18	&	5.29	&	767	&	1	&	Transit	\\
K2-24c	&	5625	&	9.18	&	7.34	&	645	&	2	&	Transit	\\
K2-25b	&	3180	&	10.44	&	3.36	&	521	&	2	&	Transit	\\
K2-260b	&	6367	&	11.09	&	17.03	&	2001	&	2	&	Eclipse	\\
K2-261b	&	5537	&	8.89	&	9.33	&	1089	&	2	&	Transit	\\
K2-266b	&	4285	&	8.9	&	3.23	&	1545	&	2	&	Transit	\\
K2-266d	&	4285	&	8.9	&	2.86	&	584	&	1	&	Transit	\\
K2-280b	&	5742	&	10.76	&	7.51	&	830	&	1	&	Transit	\\
K2-287b	&	5695	&	9.19	&	9.29	&	836	&	2	&	Transit	\\
K2-289b	&	5529	&	10.64	&	8.91	&	841	&	1	&	Transit	\\
K2-29b	&	5358	&	10.06	&	13.06	&	1193	&	2	&	Eclipse	\\
K2-3b	&	3896	&	8.56	&	2.13	&	552	&	1	&	Transit	\\
K2-30b	&	5425	&	11.09	&	11.4	&	1114	&	2	&	Eclipse	\\
K2-31b	&	5280	&	8.87	&	11.63	&	1547	&	3	&	Eclipse	\\
K2-32b	&	5275	&	9.82	&	5.03	&	840	&	2	&	Transit	\\
K2-33b	&	3540	&	10.03	&	4.94	&	801	&	1	&	Transit	\\
K2-34b	&	6071	&	10.19	&	13.66	&	1740	&	2	&	Eclipse	\\
K2-52b	&	7147	&	11.85	&	17.62	&	2242	&	2	&	Eclipse	\\
K2-55b	&	4456	&	10.47	&	3.74	&	933	&	1	&	Transit	\\
K2-99b	&	6217	&	9.72	&	11.48	&	1244	&	1	&	Transit	\\
KELT-1b	&	6518	&	9.44	&	12.18	&	2482	&	3	&	Eclipse	\\
KELT-10b	&	5948	&	9.34	&	15.35	&	1408	&	3	&	Eclipse	\\
KELT-11b	&	5375	&	6.12	&	14.81	&	1741	&	3	&	Transit	\\
KELT-12b	&	6279	&	9.36	&	19.53	&	1840	&	3	&	Eclipse	\\
KELT-14b	&	5720	&	9.42	&	19.13	&	2005	&	3	&	Eclipse	\\
KELT-15b	&	6003	&	9.85	&	19.09	&	1676	&	3	&	Eclipse	\\
KELT-16b	&	6236	&	10.64	&	15.53	&	2509	&	3	&	Eclipse	\\
KELT-18b	&	6670	&	9.21	&	17.23	&	2132	&	3	&	Eclipse	\\
KELT-2Ab	&	6327	&	7.35	&	14.81	&	1799	&	3	&	Eclipse	\\
KELT-3b	&	6304	&	8.66	&	17.12	&	1859	&	3	&	Eclipse	\\
KELT-4Ab	&	6206	&	8.69	&	18.64	&	1863	&	3	&	Eclipse	\\
KELT-6b	&	6102	&	9.08	&	14.27	&	1343	&	2	&	Eclipse	\\
KELT-7b	&	6768	&	7.54	&	17.56	&	2088	&	3	&	Eclipse	\\
KELT-8b	&	5754	&	9.18	&	17.78	&	1714	&	3	&	Eclipse	\\
KOI-12b	&	6820	&	10.23	&	16.9	&	1101	&	2	&	Eclipse	\\
KOI-94d	&	6182	&	10.93	&	11.03	&	915	&	1	&	Transit	\\
KPS-1b	&	5165	&	10.93	&	11.3	&	1482	&	2	&	Eclipse	\\
Kepler-12b	&	5947	&	12.07	&	19.25	&	1512	&	2	&	Transit	\\
Kepler-138b	&	3841	&	9.51	&	0.52	&	490	&	1	&	Transit	\\
Kepler-138d	&	3841	&	9.51	&	1.19	&	374	&	1	&	Transit	\\
Kepler-1514b	&	6251	&	10.69	&	11.58	&	417	&	1	&	Transit	\\
Kepler-16b	&	4450	&	9	&	8.27	&	234	&	2	&	Transit	\\
Kepler-18d	&	5345	&	11.76	&	6.84	&	811	&	1	&	Transit	\\
Kepler-396c	&	5384	&	10.28	&	5.19	&	505	&	1	&	Transit	\\
Kepler-422b	&	5972	&	12.04	&	12.62	&	1150	&	1	&	Transit	\\
Kepler-444b	&	5046	&	6.7	&	0.4	&	1052	&	1	&	Transit	\\
Kepler-444c	&	5046	&	6.7	&	0.48	&	973	&	2	&	Transit	\\
Kepler-444d	&	5046	&	6.7	&	0.52	&	878	&	1	&	Transit	\\
Kepler-444e	&	5046	&	6.7	&	0.54	&	815	&	1	&	Transit	\\
Kepler-447b	&	5493	&	10.81	&	18.11	&	1001	&	2	&	Eclipse	\\
Kepler-5b	&	6297	&	11.77	&	15.65	&	1847	&	2	&	Eclipse	\\
Kepler-6b	&	5647	&	11.63	&	14.31	&	1537	&	2	&	Eclipse	\\
Kepler-7b	&	5933	&	11.54	&	17.8	&	1643	&	2	&	Eclipse	\\
Kepler-76b	&	6409	&	12.09	&	14.92	&	2178	&	2	&	Eclipse	\\
Kepler-854b	&	6179	&	12.05	&	16.37	&	1854	&	2	&	Eclipse	\\
Kepler-91b	&	4550	&	10.14	&	15	&	2089	&	1	&	Eclipse	\\
LHS1140b	&	3216	&	8.82	&	1.69	&	253	&	1	&	Transit	\\
LHS1140c	&	3216	&	8.82	&	1.25	&	473	&	2	&	Transit	\\
NGTS-2b	&	6478	&	9.8	&	17.5	&	1659	&	2	&	Eclipse	\\
PH2b	&	5629	&	11.12	&	9.91	&	325	&	1	&	Transit	\\
Qatar-1b	&	5013	&	10.41	&	12.54	&	1447	&	2	&	Eclipse	\\
Qatar-2b	&	4645	&	10.62	&	13.76	&	1380	&	3	&	Eclipse	\\
Qatar-3b	&	6007	&	11.22	&	12.03	&	1716	&	1	&	Eclipse	\\
Qatar-4b	&	5215	&	11.52	&	12.45	&	1416	&	2	&	Eclipse	\\
Qatar-5b	&	5747	&	10.96	&	12.15	&	1450	&	2	&	Eclipse	\\
TRAPPIST-1b	&	2559	&	10.3	&	1.06	&	442	&	2	&	Transit	\\
TRAPPIST-1c	&	2559	&	10.3	&	1.03	&	377	&	2	&	Transit	\\
TRAPPIST-1d	&	2559	&	10.3	&	0.76	&	318	&	2	&	Transit	\\
TRAPPIST-1e	&	2559	&	10.3	&	0.9	&	277	&	2	&	Transit	\\
TRAPPIST-1f	&	2559	&	10.3	&	1.02	&	242	&	2	&	Transit	\\
TRAPPIST-1g	&	2559	&	10.3	&	1.11	&	219	&	2	&	Transit	\\
TRAPPIST-1h	&	2559	&	10.3	&	0.74	&	191	&	2	&	Transit	\\
TrES-1b	&	5230	&	9.82	&	12.4	&	1167	&	2	&	Eclipse	\\
TrES-2b	&	5850	&	9.85	&	14.92	&	1533	&	2	&	Eclipse	\\
TrES-3b	&	5650	&	10.61	&	14.66	&	1680	&	3	&	Eclipse	\\
TrES-4b	&	6200	&	10.33	&	17.67	&	1821	&	2	&	Eclipse	\\
TrES-5b	&	5171	&	11.59	&	13.1	&	1517	&	2	&	Eclipse	\\
WASP-1b	&	6304	&	10.28	&	16.27	&	1910	&	2	&	Eclipse	\\
WASP-10b	&	4675	&	9.98	&	11.85	&	993	&	3	&	Transit	\\
WASP-100b	&	6900	&	9.67	&	14.59	&	2245	&	3	&	Eclipse	\\
WASP-101b	&	6380	&	9.07	&	15.69	&	1588	&	3	&	Eclipse	\\
WASP-103b	&	6110	&	10.77	&	16.77	&	2565	&	3	&	Eclipse	\\
WASP-104b	&	5475	&	9.88	&	12.48	&	1547	&	3	&	Eclipse	\\
WASP-106b	&	6055	&	10.16	&	11.19	&	1176	&	2	&	Eclipse	\\
WASP-107b	&	4430	&	8.64	&	10.31	&	754	&	3	&	Transit	\\
WASP-11b	&	4800	&	9.42	&	12.18	&	971	&	2	&	Eclipse	\\
WASP-113b	&	5890	&	10.31	&	15.46	&	1519	&	2	&	Eclipse	\\
WASP-114b	&	5940	&	13.17	&	14.69	&	2074	&	2	&	Eclipse	\\
WASP-117b	&	6040	&	8.78	&	11.63	&	1045	&	2	&	Transit	\\
WASP-118b	&	6410	&	9.79	&	15.8	&	1765	&	2	&	Eclipse	\\
WASP-119b	&	5650	&	10.54	&	15.36	&	1602	&	2	&	Eclipse	\\
WASP-12b	&	6300	&	10.19	&	19.97	&	2638	&	3	&	Eclipse	\\
WASP-120b	&	6450	&	9.88	&	16.16	&	1918	&	2	&	Eclipse	\\
WASP-121b	&	6459	&	9.37	&	20.47	&	2413	&	3	&	Eclipse	\\
WASP-123b	&	5740	&	10.71	&	14.46	&	1550	&	3	&	Eclipse	\\
WASP-124b	&	6050	&	11.31	&	13.61	&	1420	&	2	&	Eclipse	\\
WASP-126b	&	5800	&	9.6	&	10.53	&	1520	&	2	&	Eclipse	\\
WASP-127b	&	5620	&	8.64	&	15.03	&	1431	&	3	&	Transit	\\
WASP-129b	&	5900	&	10.41	&	10.21	&	1101	&	2	&	Eclipse	\\
WASP-13b	&	5950	&	9.12	&	13.39	&	1588	&	3	&	Transit	\\
WASP-130b	&	5625	&	9.46	&	9.77	&	854	&	1	&	Eclipse	\\
WASP-131b	&	6030	&	8.57	&	13.39	&	1491	&	2	&	Transit	\\
WASP-132b	&	4775	&	9.67	&	9.55	&	781	&	2	&	Transit	\\
WASP-133b	&	5700	&	11.18	&	13.28	&	1815	&	2	&	Eclipse	\\
WASP-135b	&	5675	&	11.04	&	14.27	&	1757	&	2	&	Eclipse	\\
WASP-136b	&	6260	&	8.8	&	15.14	&	1786	&	3	&	Eclipse	\\
WASP-138b	&	6272	&	10.49	&	11.96	&	1622	&	2	&	Eclipse	\\
WASP-139b	&	5310	&	10.47	&	8.78	&	938	&	2	&	Transit	\\
WASP-14b	&	6475	&	8.62	&	15.14	&	1903	&	3	&	Eclipse	\\
WASP-140b	&	5260	&	9.16	&	15.8	&	1346	&	3	&	Eclipse	\\
WASP-141b	&	5900	&	11.19	&	13.28	&	1573	&	2	&	Eclipse	\\
WASP-142b	&	6010	&	11.44	&	16.79	&	2035	&	2	&	Eclipse	\\
WASP-144b	&	5200	&	10.9	&	9.33	&	1298	&	1	&	Eclipse	\\
WASP-145Ab	&	4900	&	9.19	&	9.88	&	1233	&	2	&	Eclipse	\\
WASP-147b	&	5702	&	10.86	&	12.24	&	1435	&	2	&	Transit	\\
WASP-15b	&	6300	&	9.69	&	15.47	&	1691	&	2	&	Eclipse	\\
WASP-151b	&	5871	&	11.19	&	12.4	&	1318	&	2	&	Transit	\\
WASP-153b	&	5914	&	11.05	&	17.01	&	1748	&	2	&	Eclipse	\\
WASP-156b	&	4910	&	9.34	&	5.6	&	992	&	1	&	Transit	\\
WASP-157b	&	5772	&	10.76	&	10.95	&	1336	&	2	&	Transit	\\
WASP-158b	&	6350	&	10.88	&	11.74	&	1623	&	2	&	Eclipse	\\
WASP-159b	&	6120	&	11.02	&	15.14	&	1889	&	2	&	Eclipse	\\
WASP-16b	&	5700	&	9.59	&	13.39	&	1334	&	2	&	Eclipse	\\
WASP-160Bb	&	5298	&	11.06	&	11.96	&	1145	&	2	&	Transit	\\
WASP-162b	&	5300	&	10.47	&	10.97	&	933	&	2	&	Transit	\\
WASP-164b	&	5806	&	10.96	&	12.38	&	1643	&	2	&	Eclipse	\\
WASP-165b	&	5599	&	11.02	&	13.83	&	1662	&	2	&	Eclipse	\\
WASP-167b	&	7000	&	9.76	&	17.34	&	2416	&	3	&	Eclipse	\\
WASP-168b	&	6000	&	10.44	&	16.46	&	1374	&	2	&	Transit	\\
WASP-17b	&	6550	&	10.22	&	20.52	&	1583	&	2	&	Eclipse	\\
WASP-172b	&	6900	&	10.13	&	17.23	&	1784	&	2	&	Eclipse	\\
WASP-173Ab	&	5800	&	10	&	13.17	&	1916	&	3	&	Eclipse	\\
WASP-174b	&	6400	&	10.58	&	14.27	&	1528	&	2	&	Eclipse	\\
WASP-18b	&	6431	&	8.13	&	13.17	&	2466	&	3	&	Transit	\\
WASP-19b	&	5568	&	10.48	&	15.27	&	2161	&	3	&	Eclipse	\\
WASP-2b	&	5200	&	9.63	&	11.75	&	1326	&	2	&	Eclipse	\\
WASP-20b	&	5940	&	9.39	&	16.04	&	1410	&	3	&	Transit	\\
WASP-21b	&	5800	&	9.98	&	11.74	&	1366	&	2	&	Eclipse	\\
WASP-22b	&	6000	&	10.32	&	13.5	&	1452	&	2	&	Eclipse	\\
WASP-23b	&	5150	&	10.45	&	10.56	&	1158	&	2	&	Eclipse	\\
WASP-24b	&	6075	&	10.15	&	15.14	&	1810	&	2	&	Eclipse	\\
WASP-25b	&	5750	&	10.17	&	11.74	&	1246	&	2	&	Eclipse	\\
WASP-26b	&	6034	&	9.69	&	13.28	&	1718	&	2	&	Eclipse	\\
WASP-28b	&	6150	&	10.73	&	13.31	&	1499	&	2	&	Eclipse	\\
WASP-29b	&	4800	&	8.78	&	8.45	&	996	&	2	&	Transit	\\
WASP-3b	&	6140	&	9.36	&	15.58	&	1717	&	3	&	Eclipse	\\
WASP-31b	&	6302	&	10.65	&	17	&	1610	&	2	&	Eclipse	\\
WASP-32b	&	6140	&	10.16	&	10.53	&	1596	&	2	&	Eclipse	\\
WASP-34b	&	5700	&	8.79	&	10.97	&	1185	&	2	&	Eclipse	\\
WASP-35b	&	5990	&	9.52	&	14.27	&	1483	&	3	&	Eclipse	\\
WASP-36b	&	5959	&	11.29	&	14.56	&	1778	&	2	&	Eclipse	\\
WASP-37b	&	5800	&	11.09	&	12.73	&	1353	&	2	&	Eclipse	\\
WASP-38b	&	6180	&	8	&	13.5	&	1285	&	3	&	Transit	\\
WASP-39b	&	5400	&	10.2	&	13.94	&	1146	&	2	&	Transit	\\
WASP-4b	&	5500	&	10.75	&	14.59	&	1708	&	3	&	Eclipse	\\
WASP-41b	&	5545	&	9.68	&	12.07	&	1271	&	2	&	Eclipse	\\
WASP-42b	&	5315	&	10.03	&	12.31	&	1044	&	2	&	Eclipse	\\
WASP-43b	&	4400	&	9.27	&	10.21	&	1409	&	3	&	Eclipse	\\
WASP-44b	&	5410	&	11.34	&	12.51	&	1381	&	2	&	Eclipse	\\
WASP-45b	&	5140	&	10.29	&	12.51	&	1224	&	2	&	Eclipse	\\
WASP-46b	&	5600	&	11.4	&	12.88	&	1676	&	2	&	Eclipse	\\
WASP-47b	&	5552	&	10.19	&	12.37	&	1291	&	2	&	Eclipse	\\
WASP-48b	&	5920	&	10.37	&	16.46	&	2078	&	2	&	Eclipse	\\
WASP-49b	&	5600	&	9.75	&	12.18	&	1401	&	2	&	Eclipse	\\
WASP-5b	&	5700	&	10.6	&	11.93	&	1745	&	2	&	Eclipse	\\
WASP-50b	&	5400	&	9.97	&	13.39	&	1422	&	2	&	Eclipse	\\
WASP-52b	&	5000	&	10.09	&	13.94	&	1330	&	3	&	Eclipse	\\
WASP-53b	&	4953	&	10.39	&	11.79	&	1079	&	2	&	Eclipse	\\
WASP-54b	&	6100	&	9.04	&	17.34	&	1822	&	3	&	Eclipse	\\
WASP-55b	&	5900	&	10.4	&	14.59	&	1296	&	2	&	Eclipse	\\
WASP-56b	&	5600	&	10.53	&	10.31	&	1230	&	2	&	Eclipse	\\
WASP-57b	&	5600	&	11.24	&	11.52	&	1371	&	2	&	Eclipse	\\
WASP-58b	&	5800	&	10.28	&	15.69	&	1306	&	2	&	Eclipse	\\
WASP-6b	&	5450	&	10.32	&	11.3	&	1210	&	2	&	Eclipse	\\
WASP-60b	&	5900	&	10.58	&	9.66	&	1343	&	1	&	Eclipse	\\
WASP-61b	&	6250	&	11.01	&	15.47	&	1583	&	2	&	Eclipse	\\
WASP-62b	&	6230	&	8.94	&	14.48	&	1459	&	3	&	Eclipse	\\
WASP-63b	&	5550	&	9.39	&	15.47	&	1565	&	2	&	Eclipse	\\
WASP-64b	&	5400	&	10.96	&	13.95	&	1685	&	2	&	Eclipse	\\
WASP-65b	&	5600	&	10.35	&	12.2	&	1518	&	2	&	Eclipse	\\
WASP-66b	&	6600	&	10.45	&	15.36	&	1841	&	2	&	Eclipse	\\
WASP-67b	&	5200	&	10.13	&	12.62	&	1049	&	2	&	Eclipse	\\
WASP-68b	&	5910	&	8.95	&	14.48	&	1516	&	2	&	Eclipse	\\
WASP-69b	&	4700	&	7.46	&	12.18	&	982	&	3	&	Transit	\\
WASP-7b	&	6400	&	8.4	&	14.59	&	1518	&	3	&	Eclipse	\\
WASP-70Ab	&	5763	&	9.58	&	12.77	&	1424	&	2	&	Eclipse	\\
WASP-71b	&	6050	&	9.32	&	12.95	&	2111	&	2	&	Eclipse	\\
WASP-72b	&	6250	&	9.62	&	14.16	&	2104	&	2	&	Eclipse	\\
WASP-73b	&	6030	&	9.03	&	15.58	&	1820	&	2	&	Eclipse	\\
WASP-74b	&	5990	&	8.22	&	14.92	&	1959	&	3	&	Eclipse	\\
WASP-75b	&	6100	&	10.06	&	13.94	&	1743	&	2	&	Eclipse	\\
WASP-76b	&	6250	&	8.24	&	20.08	&	2232	&	3	&	Transit	\\
WASP-77Ab	&	5365	&	8.4	&	15.14	&	1667	&	3	&	Eclipse	\\
WASP-78b	&	6100	&	11.01	&	21.18	&	2345	&	3	&	Eclipse	\\
WASP-79b	&	6600	&	9.06	&	18.33	&	1798	&	3	&	Eclipse	\\
WASP-8b	&	5600	&	8.09	&	12.4	&	948	&	3	&	Transit	\\
WASP-80b	&	4143	&	8.35	&	10.96	&	845	&	2	&	Eclipse	\\
WASP-81b	&	5870	&	10.89	&	15.68	&	1656	&	2	&	Eclipse	\\
WASP-82b	&	6480	&	8.76	&	17.78	&	2225	&	3	&	Eclipse	\\
WASP-83b	&	5510	&	10.39	&	11.41	&	1146	&	2	&	Transit	\\
WASP-84b	&	5314	&	8.86	&	10.34	&	817	&	2	&	Eclipse	\\
WASP-85Ab	&	6112	&	8.73	&	13.61	&	1487	&	3	&	Eclipse	\\
WASP-88b	&	6430	&	10.32	&	16.02	&	1802	&	2	&	Eclipse	\\
WASP-89b	&	5130	&	11	&	11.41	&	1149	&	2	&	Transit	\\
WASP-90b	&	6430	&	10.25	&	17.89	&	1881	&	2	&	Eclipse	\\
WASP-92b	&	6280	&	11.52	&	16.03	&	1921	&	2	&	Eclipse	\\
WASP-93b	&	6700	&	9.94	&	17.52	&	1986	&	3	&	Eclipse	\\
WASP-94Ab	&	6170	&	8.87	&	17.34	&	1536	&	3	&	Eclipse	\\
WASP-95b	&	5830	&	8.56	&	13.5	&	1649	&	3	&	Eclipse	\\
WASP-96b	&	5540	&	10.91	&	13.17	&	1314	&	2	&	Eclipse	\\
WASP-97b	&	5640	&	9.03	&	12.51	&	1576	&	3	&	Eclipse	\\
WASP-98b	&	5473	&	11.28	&	12.55	&	1197	&	2	&	Eclipse	\\
WASP-99b	&	6150	&	8.09	&	11.19	&	1501	&	2	&	Transit	\\
Wolf503b	&	4716	&	7.62	&	1.99	&	808	&	1	&	Transit	\\
XO-1b	&	5750	&	9.53	&	12.51	&	1229	&	2	&	Eclipse	\\
XO-2Nb	&	5762	&	9.31	&	10.9	&	1474	&	2	&	Eclipse	\\
XO-3b	&	6429	&	8.79	&	15.47	&	2091	&	3	&	Eclipse	\\
XO-4b	&	6397	&	9.41	&	13.72	&	1666	&	2	&	Eclipse	\\
XO-5b	&	5430	&	10.34	&	12.51	&	1254	&	2	&	Eclipse	\\
XO-6b	&	6720	&	9.25	&	22.71	&	1983	&	3	&	Eclipse	\\
\end{longtable}

\begin{table}[h!]
    \caption{Planets in the NASA Exoplanet Archive that have been detected by TESS, all of which are found to be suitable for study with Ariel.}
    \centering
    \begin{tabular}{ccccccc}
    \hline
      & \multicolumn{2}{l}{Star Properties} & \multicolumn{2}{l}{Planet Properties} & Maximum & Preferred \\
 Planet Name & Eff Temp [K] & K Magnitude &  Radius [R$_\oplus$] & Equil Temp [K] & Tier & Method \\ \hline 
GJ 143 b	&	4975	&	5.375	&	6.05	&	508	&	3 &	Transit	\\
HD 1397 b	&	5521	&	5.988	&	11.26	&	1258	& 2 &	Eclipse	\\
HD 202772 A b	&	6272	&	7.149	&	16.95	&	2181	& 3 &	Eclipse	\\
HD 219666 b	&	5527	&	8.158	&	4.61	&	1101	& 2 &	Transit	\\
HD 23472 b	&	4900	&	7.207	&	4.39	&	633	& 2 &	Transit	\\
HD 23472 c	&	4900	&	7.207	&	4.27	&	533	& 2 &	Transit	\\
LHS 3844 b	&	3036	&	9.145	&	1.27	&	827	& 2 &	Transit	\\
TOI 172 b	&	5645	&	9.722	&	10.59	&	1229	&2 &	Transit	\\
pi Men c	&	6037	&	4.241	&	2.00	&	1193	&2 &	Transit	\\ \hline
    \end{tabular}
    
    \label{TESS}
\end{table}

\newpage
\bibliography{main}

\end{document}